\documentclass[A4paper]{article}

\usepackage{amsmath,amsthm,amsfonts,amssymb}
\usepackage{upgreek}
\usepackage{euscript}
\usepackage[scale=0.7]{geometry}
\usepackage{authblk}
\usepackage{graphicx} 
\graphicspath{{./figure/}}

\newcommand{\R}{\mathbb{R}}
\newcommand{\Bc}{\mathcal{B}}
\newcommand{\Bv}{\mathcal{V}}
\newcommand{\Lc}{\mathcal{L}}
\newcommand{\cvfb}{\mathbf{C1}}
\newcommand{\cvff}{\mathbf{C2}}
\newcommand{\cvfv}{\mathbf{C2'}}

\newcommand{\ff}{\mathfrak{f}}
\newcommand{\hf}{\mathfrak{h}}

\newtheorem{remark}{Remark}

\newtheorem{definition}{Definition}
\newtheorem{theorem}{Theorem}

\newtheorem{proposition}{Proposition}
\newtheorem{example}{Example}

\title{Virtual Control Contraction Metrics: Convex Nonlinear Feedback Design via Behavioral Embedding}

\author[1]{Ruigang Wang}

\author[2,3]{Roland T\'{o}th}

\author[2]{Patrick J.W. Koelwijn}

\author[1]{Ian R. Manchester}

\affil[1]{Australian Centre for Robotics, School of Aerospace, Mechanical and Mechatronic Engineering, The University of Sydney, NSW 2006, Australia}
\affil[2]{Department of Electrical Engineering, Eindhoven University of Technology, Eindhoven, The Netherlands}
\affil[3]{Systems and Control Lab, Institute for Computer Science and Control, Budapest, Hungary}

\begin{document}

\maketitle

\begin{abstract}
    This paper presents a systematic approach to nonlinear state-feedback control design that has three main advantages: (i) it ensures exponential stability and $ \Lc_2 $-gain performance with respect to a user-defined set of reference trajectories, and (ii) it provides constructive conditions based on convex optimization and a path-integral-based control realization, and (iii) it is less restrictive than previous similar approaches. In the proposed approach, first a virtual representation of the nonlinear dynamics is constructed for which a behavioral (parameter-varying) embedding is generated. Then, by introducing a virtual control contraction metric, a convex control synthesis formulation is derived. Finally, a control realization with a virtual reference generator is computed, which is guaranteed to achieve exponential stability and $ \Lc_2 $-gain performance for all trajectories of the targeted reference behavior. We show that the proposed methodology is a unified generalization of the two distinct categories of linear-parameter-varying (LPV) state-feedback control approaches: global and local methods. Moreover, it provides rigorous stability and performance guarantees as a method for nonlinear tracking control, while such properties are not guaranteed for tracking control using standard LPV approaches.

\end{abstract}

\section{Introduction}

For linear time-invariant (LTI) systems, essentially all reasonable definitions of stability coincide, and if a particular solution (such as the zero solution)  is locally stable, then all solutions are globally stable. This favorable property extends to \textit{stabilization}, and furthermore there are many well-established methods for computing stabilizing feedback controllers.

For nonlinear systems, however, the picture is more nuanced: distinct and practically-motivated notions of stability are not necessarily equivalent. For example, stability of a particular set-point (equilibrium) does not imply stability of all set-points, which in turn does not imply stability of all non-equilibrium trajectories. Furthermore, even in the full-state-feedback case, the computation of stabilizing feedback controllers is an on-going research topic.

In this paper, we use the term \textit{regulation} to denote stabilization of a particular set-point, usually the origin under suitable choice of coordinates. For the regulation problem, the concept of a \emph{control Lyapunov function} (CLF) \cite{Sontag:1983,Artstein:1983} play an important role. Given a CLF, a simple but universal construction yields stabilizing controllers \cite{Sontag:1989}. For certain classes of nonlinear systems, CLFs can be constructed based on energy \cite{Slotine:1991}, or back-stepping techniques \cite{Krstic:1995}. For more general systems, it is desirable to construct a CLF via optimization. For linear systems, the CLF criteria can be converted into a convex \emph{linear matrix inequality} (LMI) \cite{Boyd:1994}, but for nonlinear systems, the set of CLFs for a particular system is not necessarily convex or even connected \cite{Rantzer:2001}. Certain dual notions of a CLF lead to convexity of synthesis \cite{Rantzer:2001,Lasserre:2008}, but these can only imply almost-global stabilization, and are difficult to extend to disturbance rejection or robust stabilization.

In contrast, for the much stronger problem of \textit{universal stabilization} -- i.e., global exponential stabilization of \textit{all} trajectories -- the concept of a \textit{control contraction metric} (CCM) leads to convex conditions analogous to those for the LTI case \cite{Manchester:2017}, and has been extended to disturbance rejection and robust stabilization \cite{Manchester:2018}. This concept builds on contraction analysis \cite{Lohmiller:1998}, extending it to address constructive control design. The main idea of contraction analysis is that local stability of \textit{all} trajectories of a nonlinear system implies global stability of all trajectories. Hence, stability can be addressed via analysis of an infinite family of linear systems (the local linearizations along trajectories) and is decoupled from the specification of particular trajectories. To establish stability of an LTI system or to stabilize one, it is sufficient to search via semidefinite programming for a quadratic Lyapunov function or CLF \cite{Boyd:1994}. In CCM synthesis, this is replaced by a smoothly state-varying quadratic form that measures infinitesimal distances, i.e. a Riemannian metric. The resulting search is still convex, and is defined via state-dependent \textit{point-wise} LMIs.

Nevertheless, for some systems the requirements for existence of a CCM may be too stringent. For example, in a \textit{set-point tracking} problem it may only be desired to globally stabilize a particular family of set-points: a problem that is stronger than regulation, but weaker than universal stabilization \cite{Hines:2011,Simpson-Porco:2019}. Motivated by such cases, in this paper we propose methods for achieving global exponential stability of all trajectories from a user-specified reference set $\Bc^*$. We call this problem \emph{$\Bc^*$-universal stabilization}, which covers a wide range of stabilization problems from regulation to tracking. 

Linear parameter-varying (LPV) control is a widely-applied framework that also extends LTI design techniques to nonlinear systems \cite{Rugh:2000,Hoffmann:2015}. There are two distinct categories of LPV state-feedback control approaches: local and global methods. The local LPV (a.k.a. LPV gain scheduling) approach is based on the widely-applied idea of gain scheduling \cite{Rugh:1991}. It contains three stages \cite{Rugh:2000}: linearize around a family of operating points (modeling), design controllers for those points (synthesis), and then interpolate in some way (realization). Although successfully applied in many settings, it is well known that ``hidden couplings'' between system dynamics and parameter variations can lead to closed-loop instability. Previous work to analyse this problem has led to an approach that is similar to contraction analysis, interpreting local linearization as the G\^{a}teaux derivative of a nonlinear operator \cite{Fromion:2003}.

The global LPV approach uses the same synthesis techniques as local methods, but different LPV modeling and realization methods. In this approach, the nonlinear system is modeled as an LPV system by choosing the scheduling variable as a function of state and input. This scheduling variable is treated as a ``free'' (external independent) variable throughout the synthesis step. However, it becomes in general\footnote{Except in cases where the scheduling variable is truly independent input to the system, e.g., outside temperature.} an internal variable for the control realization, resulting in an inherent conservatism that is seen as a trade-off for a convex synthesis procedure. Closed-loop stability and performance with respect to a particular set-point is guaranteed by the so-called \emph{behavior (parameter-varying) embedding principle}, i.e., any solution of the nonlinear system is also a solution of the LPV system. However, this approach can fail to guarantee asymptotic convergence in set-point tracking \cite{Scorletti:2015,Koelewijn:2020}.

In this paper, we propose a systematic approach to design nonlinear state-feedback controllers for $\Bc^*$-universal stabilization based on behavioral embedding. In this approach, we first generate behavior embeddings via the concept of a \emph{virtual system} \cite{Jouffroy:2004,Wang:2005}. The main idea of virtual system is that a nonlinear system --- which is not itself contracting --- may still have  stability properties that can be established via construction of an auxiliary observer-like system which \textit{is} contracting. This contracting virtual system takes the original state as a ``free'' parameter, and thus naturally induces a nonlinear behavior embedding of the original system, i.e., any solution of the original system is also a particular solution of the virtual system. 

By extending this virtual contraction concept, a $\Bc^*$-universal stabilizing controller can be constructed if the virtual system satisfies two conditions: $\cvfb$) the virtual system is universally stabilizable, and $\cvff$) any state reference from $\Bc^*$ is admissible to the virtual system.  For the virtual feedback design, we provide constructive conditions based on parametric LMIs via the new concept of \emph{virtual control contraction metrics} (VCCMs), which extends the CCM approach \cite{Manchester:2017} to weak stabilization problems. Moreover, since part of the system nonlinearities can be hidden into the ``free'' parameter, the synthesis problem becomes much easier compared to the CCM approach. For Condition $\cvff$, we use several examples to show that it may lead to closed-loop instability for tracking tasks if such condition does not hold. However, this condition is quite stringent as it does not hold for many virtual systems. We present a relaxed condition by introducing the concept of \emph{virtual reference generators} (VRGs), which produces a virtual reference that temporarily deviates from, but then exponentially converges to the original reference. 

We also show that the proposed VCCM approach provide a unified framework for the two distinct categories of LPV control approaches: local and global methods. Firstly, the virtual system can be taken as a generalization of the concept of \emph{global LPV embedding} to \emph{nonlinear embedding of the behavior} and the virtual differential dynamics can be seen as an extension of local linearization at operating points to continuous linearization across the entire space. Secondly, the VCCM approach provides a path-integral-based realization, which we argue is the correct realization of gain scheduling control that has been searched for in the past. We provide numerical examples in which local and global LPV approaches result in closed-loop instability at some set-points, whereas the VCCM approach guarantees global exponential stability of all set-points. We also discuss how to use our theoretic framework to explain the loss of stability guarantees for tracking control by standard LPV methods.

\paragraph{Contributions.} This paper presents a systematic behavioral embedding based nonlinear state-feedback approach for various stabilization problems -- ranging from regulation to universal stabilization. Our work has the following contributions:
\begin{itemize}
    \item We extend the CCM approach \cite{Manchester:2017} to solve weak stabilization problems. 
    Moreover, our proposed approach enables the embedding of the original dynamics into a virtual system with a simpler structure, which can reduce the complexity of contraction based control synthesis.
        
    \item We show that existing virtual contraction based control methods \cite{Manchester:2018a,Reyes:2020} require existence of a virtual feedforward controller that has restrictions on the achievable performance and even feasibility of stabilization of the system. We relax the underlying conditions by introducing VRGs and a novel feed-forward synthesis approach that ensures implementability of dynamic $\Bc^*$-universal stabilizing controllers, see Theorem~\ref{thm:nonlinear-stabilization}.
        
    \item We show that the proposed VCCM approach provides a unified theoretic framework for two distinct LPV state-feedback control methods. Our proposed approach provides rigorous stability and performance guarantees, while such properties are not guaranteed for tracking control using standard LPV approaches.
\end{itemize}


\paragraph{Related works.} Virtual or partial contraction analysis was first introduced in \cite{Jouffroy:2004} to extend contraction analysis to weaker stability notions. A more formal treatment appeared later in \cite{Wang:2005}. The first work to use virtual systems for constructive control design was \cite{Manchester:2018a}, however this was limited to Lagrangian systems. Subsequently, a virtual-contraction based approach to control of port-Hamiltonian systems was presented in \cite{Reyes:2020}. In this paper, a parametric-LMI based computational method is proposed to remove the structural assumption. For control realization, we introduce VRGs to relax the strong requirement on virtual feedforward control design \cite{Manchester:2018a, Reyes:2020, wang:2021}. Recently, we applied the proposed method to various stabilization tasks of a control moment gyroscope \cite{wang:2021}. Compared to \cite{wang:2021}, besides introducing VRGs and integrating it into the VCCM method, we provide a comprehensive description of the proposed VCCM approach, comparison with the CCM based approach, and an extensive discussion on the connection of the VCCM approach to local and global LPV methods.

\paragraph{Notation.} $ \R $ is the set of real numbers, while $ \R_+ $ is the set of non-negative reals. We use lower-case normal letters such as $x$ to denote vector signals on $\R_+$. $ \Lc_2 $ is the space of square-integrable signals, i.e., $ x\in \Lc_2$ if $\|x\|:=\sqrt{\int_{0}^{\infty}|x(t)|^2dt}<\infty$ where $ |\cdot| $ is the Euclidean norm. $ \Lc_2^e$ is the extended $\Lc_2$ space where $\|x\|_T:=\sqrt{\int_{0}^{T}|x(t)|^2dt}<\infty$ for all $ T\in\R_+$. $ A\succ (\succeq) B $ means that the matrix $ A-B$ is positive (semi)-definite.  A Riemannian metric is a smooth matrix function $ M:\R^m\rightarrow\R^{n\times n} $ with $ M(x)\succ 0 $ for all $ x\in\R^m $.  


\section{Problem Formulation and Preliminaries} \label{sec:overview} 

\subsection{Problem formulation}
Consider nonlinear systems of the form
\begin{equation}\label{eq:system}
	\dot{x}(t)=f(x(t),u(t)),
\end{equation}
where $ x(t)\in\R^{n},\,u(t)\in\R^{m} $ are the state and control input, respectively, at time $ t\in\R_+ $, and $ f $ is a smooth function of their arguments. We define the \emph{behavior} $\Bc$ as the set of all forward-complete solutions of \eqref{eq:system}, i.e. $ \Bc=\{(x,u)\}$ with $ x:\R_+\rightarrow\R^n$ piecewise differentiable and $u:\R_+\rightarrow\R^m$ piecewise continuous satisfying \eqref{eq:system} for all $t\in \R_+$. A state trajectory $ x $ is said to be admissible to system \eqref{eq:system} if there exists an input signal $ u $ such that $ (x,u)\in\Bc $. 

Let $\Bc^*\subset \Bc$ be a (feasible) set of desired behaviors. We consider state-feedback controllers that explicitly depend on $ (x^*,u^*)\in\Bc^* $:
\begin{equation}\label{eq:controller}
	u(t)=\kappa(x(t),x^*(t),u^*(t)),
\end{equation}
where $ \kappa:\R^n\times\R^n\times\R^m\rightarrow\R^m $. By applying \eqref{eq:controller} to system \eqref{eq:system} we obtain the closed-loop system
\begin{equation}\label{eq:cl-system}
	\dot{x}(t)=f(x(t),\kappa(x(t),x^*(t),u^*(t))).
\end{equation}
The reference trajectory $ (x^*,u^*) $ is said to be  globally exponentially stabilizable if one can construct a feedback controller \eqref{eq:controller} such that for any initial condition $ x(0)\in\R^n $, a unique solution $ x $ exists for \eqref{eq:cl-system} and satisfies
\begin{equation}\label{eq:stability}
	|x(t)-x^*(t)|\leq Re^{-\lambda t}|x(0)-x^*(0)|,
\end{equation}
where rate $ \lambda>0 $ and overshoot $ R>0 $ are constants independent of initial conditions but may depend on the choice of reference trajectory. If every $(x^*,u^*)\in\Bc^*$ is globally exponentially
stabilizable, then the system is said to be
\emph{$\Bc^*$-universally exponentially stabilizable}. If $ \Bc_1^*\supset \Bc_2^*$, then $\Bc_1^*$-universal stabilization is stronger than $\Bc_2^*$-universal stabilization. The strongest case is $ \Bc^*=\Bc $, which is called universal stabilization \cite{Manchester:2017}.

We will also consider the disturbance rejection problem for the perturbed system of \eqref{eq:system}:
\begin{equation}\label{eq:system-pert}
	\dot{x}(t)=f(x(t),u(t),w(t)),\quad z(t)=h(x(t),u(t),w(t)),
\end{equation}
where $ w(t)\in\R^{p},z(t)\in\R^{q} $ are collections of external disturbances (load, noise, etc.) and performance outputs (tracking error, actuator usage, etc.), respectively. With slight abuse of notation, we use $ \Bc $ to denote the behavior of \eqref{eq:system-pert}. The set $ \Bc^* \subset \Bc $ is called a reference behavior if each $(x^*,u^*,w^*,z^*)\in\Bc^* $ satisfies $ w^*=0 $ (i.e., the nominal value of disturbance is 0). Similar to the stabilization problem, we consider state-feedback controllers of the form \eqref{eq:controller}, leading to the closed-loop system:
\begin{equation}\label{eq:sys-cl}
\begin{split}
	\dot{x}(t)=f(x(t),\kappa(x(t),x^*(t),u^*(t)),w(t)), \quad
	z(t)=h(x(t),\kappa(x(t),x^*(t),u^*(t)),w(t)).
\end{split}
\end{equation}
The above controlled system is said to have \emph{$ \Bc^* $-universal $ \Lc_2 $-gain} bound of $ \alpha $ if for each $ (x^*,u^*,w^*,z^*)\in\Bc^* $, any initial condition $ x(0) \in \R^n $ and any input $ w \in \Lc_2^e $, a unique solution $ (x,w,z) $ of \eqref{eq:sys-cl} exists and satisfies
\begin{equation}\label{eq:L2-gain}
	\|z-z^*\|_T^2\leq \alpha^2\|w-w^*\|_T^2+\beta(x(0),x^*(0)),
\end{equation}
for all $ T>0 $ and some continuous function $ \beta(x,y)\geq 0 $ with $ \beta(x,x)=0 $. 


The controller synthesis problems we address in this paper are (i), synthesizing a state-feedback controller of the form \eqref{eq:controller} for a nonlinear system of the form \eqref{eq:system} such that the resulting closed-loop system (given by \eqref{eq:cl-system}) is $\Bc^*$-universally exponentially stable, and (ii), synthesizing a state-feedback controller of the form \eqref{eq:controller} for a nonlinear system of the form \eqref{eq:system-pert} such that the resulting closed-loop system (given by \eqref{eq:sys-cl}) is $\Bc^*$-universally exponentially stable and satisfies a prescribed $ \Bc^* $-universal $ \Lc_2 $-gain bound.

In the next section, we will briefly discuss contraction and virtual contraction theory, which will be used to address these controller synthesis problems.

\subsection{Contraction analysis}   
Contraction analysis studies the convergence between arbitrary trajectories of the system, which has proven to be a useful tool for constructive design of nonlinear tracking controllers \cite{Manchester:2017,Wang:2017}. Here we summarize the main results of contraction theory, see details in \cite{Lohmiller:1998,Forni:2014} and the survey \cite{Aminzare:2014}. Consider a nonlinear time-varying system of the form:
\begin{equation}\label{eq:sys-auto}
\dot{x}(t)=f(t,x(t)),
\end{equation}
where $ x(t)\in\R^n $ is the state, and $ f $ is a smooth function. Note that time variation will appear in the virtual systems associated with time-invariant systems \eqref{eq:system} and \eqref{eq:system-pert}. System~\eqref{eq:sys-auto} is said to be \emph{contracting} if any solution pair $ (x_1,x_2) $ satisfies
\[
	|x_2(t)-x_1(t)|\leq Re^{-\lambda t}|x_2(0)-x_1(0)|,
\]
where $ \lambda, R$ are some positive constants. To analyze the contraction property, we utilize the ``extended'' system consisting of \eqref{eq:sys-auto} and its differential dynamics:
\begin{equation}\label{eq:diff-dyn}
	\dot{\delta}_{x}(t) =A(t,x)\delta_{x}(t):=\frac{\partial f(t,x)}{\partial x}\delta_{x}(t),
\end{equation}
defined along the solutions $ x $. The state $\delta_x(t)$ is the infinitesimal variation \cite{Lohmiller:1998,Aminzare:2014} (or can be seen as a tangent vector \cite{Forni:2014}) at the point $x(t)$. A uniformly-bounded Riemannian metric $ M(x,t) $, i.e. $a_1I\preceq M(x,t)\preceq a_2I$, is called a {contraction metric} for \eqref{eq:sys-auto} if 
\begin{equation}\label{eq:cond-contraction}
	\dot{M}+MA+A^\top M\preceq -2\lambda M,
\end{equation}
for all $ x\in \R^n,t\in\R_{+} $, where $ \dot{M}=\frac{\partial M}{\partial t}+\sum_{i=1}^{n}f_i\frac{\partial M}{\partial x_{i}} $. The contraction metric $ M(x,t) $ also induces a quadratic differential Lyapunov function $ V(x,t,\delta_x)=\delta_x^\top M(x,t)\delta_x $ for \eqref{eq:diff-dyn}, i.e., $ \dot{V}\leq -2\lambda V $. A central result of \cite{Lohmiller:1998} is that the existence of a contraction metric for system \eqref{eq:system} implies that it is contracting with $\lambda$ and $R=\sqrt{a_2/a_1}$.

Contraction analysis can be extended to the system with external input $ w(t)\in\R^p $ and performance output $ z(t)\in\R^q $:
\begin{equation}\label{eq:open-sys}
\dot{x}=f(t,x,w),\quad z=h(t,x,w),	
\end{equation}
whose differential dynamics is of the form:
\begin{equation}\label{eq:open-sys-diff}
	\begin{split}
		\dot{\delta}_{x} =A(t,x,w)\delta_{x}+B(t,x,w)\delta_{w}, \quad \delta_{z}=C(t,x,w)\delta_{x}+D(t,x,w)\delta_{w},
	\end{split}
\end{equation}
where $ A=\frac{\partial f}{\partial x} $, $ B=\frac{\partial f}{\partial w} $, $ C=\frac{\partial h}{\partial x} $, and $ D=\frac{\partial h}{\partial w} $. System \eqref{eq:open-sys} is said to have a \emph{differential $ \Lc_2 $-gain bound} of $ \alpha $, if for all $ T>0 $,
\begin{equation}\label{eq:diff-L2-gain}
	\|{\delta}_z\|_T^2\leq \alpha^2\|{\delta}_w\|_T^2+b(x(0),\delta_{x}(0)),
\end{equation}
where $ b(x,\delta_x)\geq 0 $ with $ b(x,0)=0 $ for all $ x $. From \cite[Th. 3.1.11]{Schaft:1999}, a sufficient, and in some cases necessary, condition for \eqref{eq:diff-L2-gain} is the existence of a differential storage function $ V(x,t,\delta_x)\geq 0 $ with $ V(x,t,0)=0 $ that verifies
\begin{equation}
V_{t_2}-V_{t_1}\leq \int_{t_1}^{t_2} \left(-\delta_{z}^\top\delta_{z}+\alpha^2\delta_{w}^\top\delta_{w}\right) \ dt,
\end{equation}
where $ V_t=V(x(t),t,\delta_{x}(t)) $. For smooth systems, the differential $ \Lc_2 $-gain bound is equivalent to the incremental $ \Lc_2 $-gain bound \cite{Fromion:2003}.

\subsection{Behavior embedding and virtual contraction analysis}
Virtual contraction analysis extends contraction theory to study convergence between behaviors (i.e., trajectory sets). Consider a time-invariant smooth autonomous system
\begin{equation}\label{eq:sys-auto-ti}
	\dot{x}=f(x),
\end{equation} 
with $ x\in\R^n $. A \emph{virtual system} is a new system of the form
\begin{equation}\label{eq:sys-auto-virtual}
	\dot{y}=\ff(y,x),
\end{equation}
with $ \ff(x,x)=f(x) $ and $\ff$ being smooth, where the virtual state $ y $ lives in a copy of the original state space $ \R^n $, and the variable $ x $, taken as an exogenous input, is the state associated with the original system \eqref{eq:sys-auto-ti}. One way to construct a virtual system is based on the factorization of the dependency of $f$ on $x$, e.g., $f(x)=x^2$ can be re-casted as $\ff(y,x)=xy$. Hence, construction of \eqref{eq:sys-auto-virtual} is non-unique. A virtual system introduces an \emph{embedding} relationship between the behavior $\Bc$ of \eqref{eq:sys-auto-ti} and the virtual behavior $\Bv$ of \eqref{eq:sys-auto-virtual}. That is, $ x \in \Bc$ is equivalent to $ (x,x)\in \Bv$. For any $x\in \Bc$, we can define a \emph{projected virtual behavior} $ \Bv_{x} =\{y \mid (y,x)\in\Bv\} $, which is the state trajectory set of \eqref{eq:sys-auto-virtual} with fixed external signal $x$. 

Associated with $ \Bv_{x} $, the virtual differential dynamics is 
\begin{equation}\label{eq:diff-dyn-virtual}
	\dot{\delta}_{y} =\frac{\partial \ff(y,x)}{\partial y}\delta_{y}.
\end{equation}
Note that $ \delta_{x}(t)=0 $ for all $t$ as $ x $ is a fixed external signal for all $ y\in \Bv_{x} $. System \eqref{eq:sys-auto-ti} is said to be \emph{virtually contracting} if there exists a metric $ M(y,x) $ such that for any $ x\in\Bc $, $ \widehat{M}(y,t)=M(y,x(t)) $ is a contraction metric for the time-varying system \eqref{eq:sys-auto-virtual}. Here, $ M(y,x) $ is also referred to as a virtual contraction metric. The existence of a virtual contraction metric implies that all virtual trajectories in $ \Bv_{x} $ converge to each other, and thus they converge to a particular solution $ x $ as $ x\in\Bv_{x}$. Furthermore, if $ \Bv_{x} $ also contains $ x^*\in\Bc^*$, we can conclude that $ |x(t)-x^*(t)| $ vanishes exponentially.

\begin{theorem}[Virtual contraction \cite{Wang:2005}]\label{thm:virtual-contraction}
	Assume that \eqref{eq:sys-auto-ti} is a virtually contracting system. A trajectory $ x\in \Bc $ exponentially converges to $ x^*\in\Bc^* $ if $ x^*\in\Bv_{x}$. 
\end{theorem} 

\section{Nonlinear Stabilization via VCCM}\label{sec:stabilization}

In this section we first present the behavioral embedding based control framework for the $ \Bc^* $-universal stabilization problem. In this framework there are two main components: virtual feedforward and feedback controllers. We give a convex formulation for designing the virtual feedback controller by introducing the new concept of virtual control contraction metrics (VCCMs). After that, we introduce the concept of virtual reference generator, which can be understood as an extension of the virtual feedforward controller. Finally, we discuss the differences between the proposed approach and the CCM-based control approach. For the sick of simplicity, the dependency of time $t$ is dropped whenever it can be clearly inferred from the context. 


\subsection{Behavior embedding based control framework}

A virtual system of \eqref{eq:system} can be represented by
\begin{equation}\label{eq:sys-virt}
	\dot{y}=\ff(y,x,v),
\end{equation}
with $ \ff(x,x,u)=f(x,u) $, where $ y(t)\in\R^{n},\, v(t)\in\R^m $ are the virtual state and input, respectively, and the variable $ x $, taken as an exogenous variable, is the state of the original system \eqref{eq:system}.
\begin{remark}
    If $\ff$ is chosen to be linear in $y$ and $v$, the virtual system \eqref{eq:sys-virt} become a conventional global LPV embedding \cite{Toth2010SpringerBook}. By allowing for more general behavior embeddings (e.g. nonlinear parameter-varying embedding), we can further reduce the conservativeness due to embedding nature while providing a similar convex control formulation as the LPV embedding. 
\end{remark}

To obtain a $\Bc^*$-universal stabilizing controller, we construct virtual system \eqref{eq:sys-virt} based on the following two conditions:
\begin{enumerate}
	\item[{$ \cvfb $}] (virtual feedback) For any admissible state trajectory $ x $ of $ \Bc $, the projected virtual behavior $ \Bv_{x}:=\{(y,v)\mid(y,x,v)\in\Bv\} $ is universally exponentially stabilizable, i.e., for any virtual reference $ (y^*,v^*)\in\Bv_{x} $ there exists a \emph{virtual feedback controller} of the form
	\begin{equation}\label{eq:virt-fb}
		v=\kappa^\mathrm{fb}(y,x,y^*,v^*),
	\end{equation}
	where $ \kappa^\mathrm{fb}:\R^n\times\R^n\times\R^n\times\R^m\rightarrow\R^m $ is a locally Lipschitz continuous map, such that \eqref{eq:sys-virt} is exponentially stabilized by \eqref{eq:virt-fb} at $ (y^*,v^*) $.

	\item[$ \cvff $] (virtual feedforward) For any admissible trajectory $ x $ of $ \Bc $ and any reference trajectory $ (x^*,u^*)\in\Bc^*\subseteq \Bc $, there exists a \emph{virtual feedforward controller} of the form
	\begin{equation}\label{eq:virt-ff}
		v^*=\kappa^\mathrm{ff}(x,x^*,u^*),		
	\end{equation}
	where $ \kappa^\mathrm{ff}:\R^n\times\R^n\times\R^m\rightarrow\R^m $ is a locally Lipschitz continuous map, such that $ u^*=\kappa^\mathrm{ff}(x^*,x^*,u^*) $ and $ (x^*,v^*)\in\Bv_{x} $ hold, i.e., any state reference from $\Bc^*$ is admissible to the virtual system.
	
\end{enumerate}

By substituting \eqref{eq:virt-ff} into \eqref{eq:virt-fb} and setting $ y^*=x^* $, system \eqref{eq:sys-virt} is globally exponentially stabilized at the reference $x^*$ via the following controller:
\begin{equation}\label{eq:virt-control}
	\begin{split}
	v=\overline{\kappa}(y,x,x^*,u^*) 
	:=\kappa^\mathrm{fb}(y,x,x^*,\kappa^\mathrm{ff}(x,x^*,u^*)).
	\end{split}
\end{equation} 
By setting $y=x$ and $v=u$, we obtain an overall controller for system \eqref{eq:system}: 
\begin{equation}\label{eq:true-control}
	u=\kappa(x,x^*,u^*):=\overline{\kappa}(x,x,x^*,u^*).
\end{equation}
Closed-loop stability with $\kappa$ is guaranteed by the following theorem.

\begin{theorem}\label{thm:vccm}
	Consider the system \eqref{eq:system} and a reference behavior $ \Bc^* $. Assume that there exists a virtual system representation \eqref{eq:sys-virt} such that Conditions $\cvfb $ and $\cvff$ hold. Then, for any $\kappa^{\mathrm{fb}} $ and $\kappa^{\mathrm{ff}} $ that satisfy $\cvfb$ and $\cvff$, the closed-loop system \eqref{eq:system} under \eqref{eq:true-control} is $ \Bc^* $-universally exponentially stable.
\end{theorem}
\begin{proof} 
	From \eqref{eq:sys-virt} and \eqref{eq:virt-control}, we can obtain a virtual closed-loop system of the form
	\begin{equation}\label{eq:virt-cl-system}
		\dot{y}=\ff(y,x,\overline{\kappa}(y,x,x^*,u^*)),
	\end{equation}
	which is contracting by Condition~$\cvfb$. Moreover, $\overline{\kappa}$ is locally Lipschitz continuous due to the definition of $\kappa^{\mathrm{fb}}$ and $\kappa^{\mathrm{ff}}$. Then, \eqref{eq:virt-cl-system} is forward complete by \cite[Thm.2]{Angeli:1999}. Let $ \Bc^\kappa$ and $\Bv^\kappa $ be the behaviors of \eqref{eq:cl-system} and \eqref{eq:virt-cl-system}, respectively. Since \eqref{eq:virt-cl-system} is a virtual system of \eqref{eq:cl-system}, we have $x\in \Bv_{x}^\kappa$ for any $x\in \Bc^\kappa$ by the behavior embedding principle. Condition~$ \cvff $ implies that $ x^*\in\Bv_{x}^\kappa $ for all $x$. Thus, Theorem~\ref{thm:vccm} is implied by Theorem~\ref{thm:virtual-contraction}.
\end{proof}

\begin{example}\label{exmp:1}
	Consider a system with state $x=[x_1\ x_2]^\top$ and nonlinear dynamics 
	\begin{equation}\label{eq:sys-1}
		\dot{x}=A(x)x+Bu:= \begin{bmatrix}
			-1 & 0 \\ -\sin x_2 & -1
			\end{bmatrix}x+\begin{bmatrix}
				1 \\ 0
			\end{bmatrix}u.
	\end{equation} 
	This system is not universally stabilizable as any trajectory starting in the plane $ x_2=0 $ will remain in this plane. However, it can be $ \Bc^* $-universally stabilized if all reference trajectories in $ \Bc^* $ satisfy $ x_{2}^*(t)=0,\, \forall t\in \R_+$. First, we choose the virtual system $ \dot{y}=A(x)y+Bv $ and then take the virtual feedback controller as
	\begin{equation}
	    v=v^*+(y_2-y_2^*)\sin x_2,
	\end{equation}
	where $(v^*,y^*)$ is the virtual reference trajectory. It is easy to verify that the above controller satisfies Condition~$\cvfb$. Moreover, Condition~$\cvff$ also holds if we take the virtual feedforward controller as $y_2^*=x_2^*=0$ and $v^*=u^*$. By the embedding principle, we obtain a $\Bc^*$-universal stabilizing controller $u=u^*+x_2\sin x_2$ for the original nonlinear system \eqref{eq:sys-1}. 
\end{example}
	
\subsection{Virtual control contraction metrics} \label{sec:VCCM:concept}

Here we present a constructive method for designing a virtual feedback controller satisfying Condition~$\cvfb$ via VCCMs. It can be viewed as an extension of the CCM approach \cite{Manchester:2017} to $\Bc^*$-universal stabilization. 

For any fixed admissible state trajectory $x$ of $\Bc$, the virtual system \eqref{eq:sys-virt} becomes a time-varying nonlinear system. Its associated differential dynamics can be given by
\begin{equation}\label{eq:sys-virt-diff}
\begin{split}
	\dot{\delta}_{y}&=A(\sigma)\delta_{y}+B(\sigma)\delta_{v},
\end{split}
\end{equation} 
where $ \sigma=(y,x,v) $, $ A=\frac{\partial \ff}{\partial y} $ and $ B=\frac{\partial \ff}{\partial v} $. 
\begin{remark}
    Note that \eqref{eq:sys-virt-diff} fulfills the properties of an LPV system (see \cite{Toth2010SpringerBook})  with $\upsigma$ as the scheduling-variable. In fact, if one takes the original nonlinear system \eqref{eq:system} as a virtual system \eqref{eq:sys-virt} (i.e. $\ff =f$), then \eqref{eq:sys-virt-diff} is a generalized local LPV embedding, see more details in Section~\ref{sec:comparison}. 
\end{remark}

We now give the definition of VCCMs as follows.
\begin{definition}
    A uniformly-bounded Riemannian metric $ M(y,x) $ is said to be a virtual control contraction metric for \eqref{eq:system}, if the following implication is true for all $ y,x,v $:
\begin{equation}\label{eq:vccm}
\begin{split}
\delta_{y}&\neq 0,\,\delta_{y}^\top M B=0 \implies \delta_{y}^\top(\dot{M}+MA+A^\top M+2\lambda M)\delta_{y}<0.
\end{split}
\end{equation}
\end{definition}
The following result is a direct application of \cite[Thm.~1]{Manchester:2017}.
\begin{proposition}\label{prop:virtual-contraction}
	Given a virtual system \eqref{eq:sys-virt} based embedding of \eqref{eq:system}, Condition~$\cvfb$ holds if there exists a VCCM.
\end{proposition}

\subsubsection{Control synthesis.}
We consider virtual systems with the following control-affine form
\begin{equation}
    \dot{y}=g(y,x)+B(y,x)v.
\end{equation}
Its differential dynamics can be written as 
\begin{equation}\label{eq:cfs-diff-dyn}
    \dot{\delta}_y=A(y,x,v)\delta_y+B(y,x)\delta_v:=\left(\frac{\partial g}{\partial y}+\sum_iv_i\frac{\partial b^i}{\partial y}\right) \delta_y + B(y,x)\delta v
\end{equation}
where $b^i$ is the $i$th column of $B(y,x)$. The VCCM-based control synthesis problem can be formulated as the following conditions on a uniformly-bounded metric $W(y,x):\R^n\times \R^n\rightarrow \R^n$ and matrix function $ Y(y,x):\R^n\times\R^n\rightarrow\R^{m\times n} $:
\begin{subequations}\label{eq:vccm-synthesis}
	\begin{gather}
		-\sum_{i=1}^{n}\dot{x}_i\frac{\partial W}{\partial x_i}+\frac{\partial g}{\partial y}W+ W \left(\frac{\partial g}{\partial y}\right)^\top +BY+Y^\top B^\top + 2\lambda W\preceq 0, \label{eq:vccm-1} \\
		\sum_{j=1}^{n}b_j^i\frac{\partial W}{\partial y_j}-\frac{\partial b^i}{\partial y}W- W\left(\frac{\partial b^i}{\partial y}\right)^\top=0,\; i=1,\ldots,n, \label{eq:vccm-2}
	\end{gather}
\end{subequations}
where $b_j^i$ is the $(i,j)$th element of $B(y,x)$. Condition \eqref{eq:vccm-1} says that the uncontrolled system is contracting in directions orthogonal to the span of the control inputs. Condition \eqref{eq:vccm-2} implies that $\dot{W}$ is independent of $v$, i.e., the vector fields $b^i$ are Killing fields for the metric $W$. Based on \eqref{eq:vccm-synthesis} one can obtained that
\begin{equation}\label{eq:vccm-strong}
	\dot{M}+M(A+BK)+(A+BK)^\top M+2\lambda M\preceq 0,
\end{equation}
where $M(y,x)=W^{-1}(y,x)$ and $K(y,x)=Y(y,x)W^{-1}(y,x)$. Since \eqref{eq:vccm-strong} implies \eqref{eq:vccm}, we have that $M(y,x)$ is a VCCM. Moreover, \eqref{eq:cfs-diff-dyn} is exponentially stable under the differential controller 
\begin{equation}\label{eq:diff-controller}
\delta_{v}=K(y,x)\delta_{y}.
\end{equation}

The formulation in \eqref{eq:vccm-synthesis} is convex in $ W,Y $, but infinite dimensional as the decision variables are sets of smooth matrix functions. There are various finite-dimensional LMI approximations to turn \eqref{eq:vccm-synthesis} into efficiently computable synthesis problems. One way is to apply an LPV synthesis technique, since the differential dynamics \eqref{eq:sys-virt-diff} vary with $\sigma$, which can be seen as the scheduling-variable. By computing a convex outer approximation (e.g. a convex polytope) of the possible signal variations $(\sigma,\dot{x})$ in $\Bv $,  \eqref{eq:vccm-synthesis} can be transformed to a finite set of LMI constraints using  LPV state-feedback synthesis techniques if $A,B$ and $Y$ are rational matrix functions in $\sigma $, and $ W $ is rational matrix function in $ y,x $ (see \cite{Hoffmann:2015} for an overview). An alternative representation for $A,B,Y$ is the gridding based approach, see \cite{hjartarson2015lpvtools} for details. Another way is to approximate the components of $ (g,B,Y,W) $ by polynomials up to a chosen order, and verifying the inequalities by the sum-of-squares relaxation \cite{Parrilo:2003}. 

\subsubsection{Control realization.}

We construct the virtual feedback realization by integrating the above differential controller \eqref{eq:diff-controller} along a particular path, i.e.
\begin{equation}\label{eq:virt-controller}
    v(t)=\kappa^\mathrm{fb}(y(t),x(t),y^*(t),v^*(t)):=v^*(t)+\int_{0}^{1}K(\gamma(s),x(t))\gamma_s(s)ds,
\end{equation}
where the path $\gamma(s): [0,1]\rightarrow \R^n$ is a (shortest path) geodesic $\gamma$ connecting $y^*(t)$ to $y(t)$ w.r.t. the metric $M(y,x)$. To be specific, $\gamma$ is computed by solving the following optimization problem
\begin{equation}\label{eq:geodesic}
	\begin{split}
	\min_{c:[0,1]\rightarrow\R^n}&\int_{0}^{1}c_s^\top M(c(s),x(t))c_s ds \quad 
	\mathrm{s.t.}\quad c(0)=y^*(t),\;c(1)=y(t),
	\end{split}
\end{equation} 
where $c_s:=dc/ds$, see \cite{Leung:2017,wang:2019} for detailed algorithms. If the metric $M$ is independent of $y$, then $\gamma$ is simply a straight line, i.e. $\gamma(s)=sy(t)+(1-s)y^*(t).$ Based on \cite{Manchester:2017}, we can conclude that \eqref{eq:virt-controller} is a virtual feedback controller satisfying Condition $\mathbf{C1}$.






\subsubsection{Case study.}\label{sec:case-study} 
\begin{figure}[!bt]
    \centering
    \includegraphics[width=0.2\linewidth]{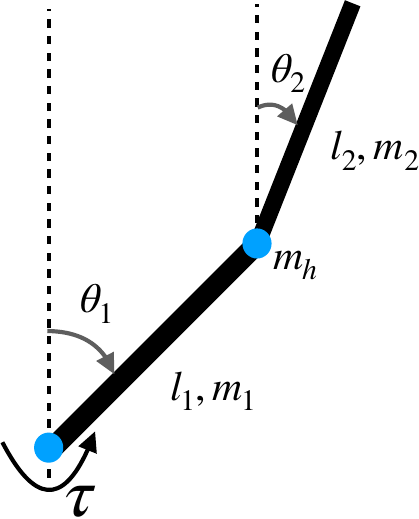}
    \caption{The pendubot system (see physical parameters in \cite{Cisneros:2019}).}\label{fig:pendubot-scheme}
\end{figure}

\begin{figure}[!bt]
    \centering
    \includegraphics[width=0.7\linewidth]{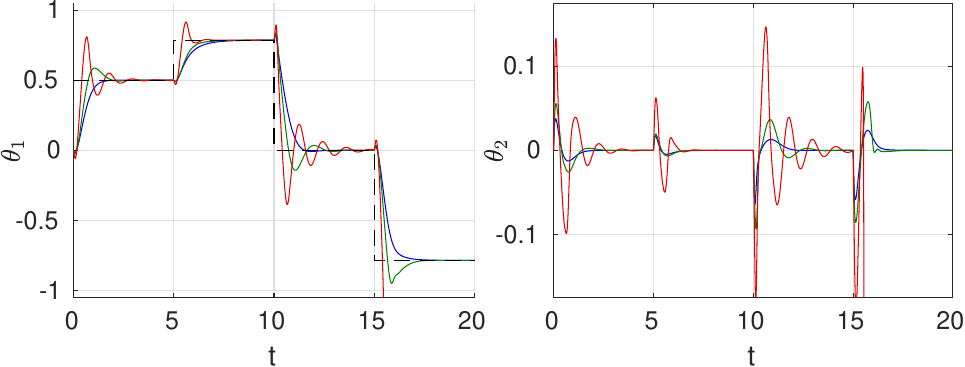}
    \caption{State responses of \eqref{eq:pendubot} with VCCM controllers based on different virtual systems: blue - \eqref{eq:vs-1}, green - \eqref{eq:vs-2}, red - \eqref{eq:vs-3}.}\label{fig:pendubot}
\end{figure}

We illustrate our approach on the pendubot system from \cite{Cisneros:2019}. The configuration, coordinates, and relative parameters of the pendubot are depicted in Fig.~\ref{fig:pendubot-scheme}. 
The pendubot dynamics is a nonlinear system of the form
\begin{equation}\label{eq:pendubot}
M(\theta)\ddot{\theta}+C(\theta,\dot{\theta})\dot{\theta}+G(\theta)=R \tau
\end{equation}
where $ \theta=[\theta_1,\theta_2]^\top$ is the angle configuration of two links and $\tau$ is the control input torque at the base joint. Here the coefficient matrices are
\[
		 \begin{split}
			M=&
			\begin{bmatrix}
				(m_1+m_2)l_1^2 & m_2l_1l_2\cos(\theta_1-\theta_2) \\
				m_2l_1l_2\cos(\theta_1-\theta_2) & m_2 l_2^2
			\end{bmatrix}, \\
			C=&
			\begin{bmatrix}
				f_{v1} & m_2l_1l_2\sin(\theta_1-\theta_2)\dot{\theta}_2 \\ -m_2l_1l_2\sin(\theta_1-\theta_2)\dot{\theta}_1 & f_{v2}
			\end{bmatrix}, \\
			G=&
			\begin{bmatrix}
				-(m_1/2+m_2+m_h)l_1g\sin\theta_1 \\
				-m_2l_2g\sin\theta_2
			\end{bmatrix},\quad R=\begin{bmatrix}
			    1 \\ 0
			\end{bmatrix}
		 \end{split}
\]
where $f_{v1}, f_{v2}$ are the friction coefficients of two joints. We consider the stabilization problem of unstable set-points over a large range, i.e. the reference $\theta^*\in:=\{(\theta_1^*,\theta_2^*)\mid |\theta_{1}^*|\leq \pi/3, \theta_2^*=0\}$. Our proposed approach takes the following steps. 
\begin{itemize}
    \item[1)] \emph{Behavior embedding via virtual system}. Here we construct three virtual systems via different treatment of nonlinearities in $M,C,G$:
	\begin{subequations}
		 \begin{align}
			 M(\theta)\ddot{\phi}+C(\phi,\dot{\phi})\dot{\phi}+G(\phi)&=R v, \label{eq:vs-1}\\
			 M(\theta)\ddot{\phi}+C(\theta,\dot{\theta})\dot{\phi}+G(\phi)&=R v, \label{eq:vs-2}\\
			 M(\theta)\ddot{\phi}+C(\theta,\dot{\theta})\dot{\phi}+H(\theta)\phi&=R v, \label{eq:vs-3}
		 \end{align}
	\end{subequations}
    where $\phi,v$ are the virtual angle and input, respectively. The matrix $H(\theta)\in \R^{2\times 2}$ is chosen such that $H(\theta)\theta=G(\theta)$ for all $\theta\in \R^2$. Note that such $H$ always exists as $G$ is differentiable. System \eqref{eq:vs-1} only hides the nonlinear terms $M(\theta)\ddot{\theta}$ of \eqref{eq:pendubot} into the linear embedding $M(\theta)\ddot{\phi}$ while keeping the rest unchanged. System \eqref{eq:vs-3} takes a full LPV embedding for all the nonlinear terms. System \eqref{eq:vs-2} sits between \eqref{eq:vs-1} and \eqref{eq:vs-3}. By introducing the virtual state $y=[\phi,\dot{\phi}]^\top$ and external parameter $x=[\theta,\dot{\theta}]^\top$, we rewrite \eqref{eq:vs-1} -- \eqref{eq:vs-3} as control-affine virtual systems \eqref{eq:cfs-diff-dyn}.
    
    \item[2)] \emph{Control synthesis}.  For each virtual system representation, we apply the gridding method to solve \eqref{eq:vccm-synthesis} with $\lambda=0.5$ to search for a constant $W\succ 0$ and a matrix function $Y(y,x)$ over the operating region $ |\theta_1|\leq \pi/3, |\theta_2|\leq \pi/18, |\dot{\theta}_1|\leq 1, |\dot{\theta}_2|\leq 1$. 
    
    \item[3)] \emph{Control realization}. Since the VCCM $M=W^{-1}$ is a constant metric, the virtual feedback controller in \eqref{eq:virt-controller} can be written as
    \begin{equation}
        v=\kappa^{\mathrm{fb}}(y,x,y^*,v^*):=v^*+\left(\int_{0}^1 K(\gamma(s), x)ds \right)(y-y^*),
    \end{equation}
    with $K(y,x)=Y(y,x)W^{-1}$ is the differential control gain and $ \gamma(s)=(1-s)y^*+sy$ is the geodesic connecting $y^*$ to $y$. Note that the above virtual feedback controller satisfies Condition~$\cvfb$ when the scheduling variable $(y,x)$ is within the operating region. For virtual feedforward design, we choose $y^*=x^*:=[\theta^*,0]^\top$ and $v^*=\kappa^{\mathrm{ff}}(x,x^*,u^*):=u^*$ where $u^*$ is the nominal input torque satisfying $Ru^*=G(\theta^*)$. Finally, by applying the embedding rule $y(t)=x(t)$ and $u(t)=v(t)$ for all $t$,  we obtain the overall controller for the \eqref{eq:pendubot} as
    \begin{equation}
        u=u^*+\left(\int_{0}^1 K(\gamma(s), x)ds \right)(x-x^*),
    \end{equation}
    where $\gamma(s)=(1-s)x^*+sx$.
    
\end{itemize}
Fig.~\ref{fig:pendubot} depicts the closed-loop responses for control design based on virtual systems \eqref{eq:vs-1} -- \eqref{eq:vs-3}. The controllers from \eqref{eq:vs-1} and \eqref{eq:vs-2} can stabilize the system at different set points while instability behavior is observed for the controller based on \eqref{eq:vs-3} at the reference point $\theta_1^*=-\pi/4$. The explanation is as follows. First, we can easily verify that the chosen feedforward controller $\kappa^{\mathrm{ff}}$ satisfies Condition~$\cvff$ under the virtual systems \eqref{eq:vs-1} and \eqref{eq:vs-2}. Thus, $\Bc^*$-universal stability follows by Theorem~\ref{thm:vccm}. Secondly, for the virtual system \eqref{eq:vs-3}, Condition~$\cvff$ requires $G(x)x^*=u^*,\forall x$,  which only holds when $\theta_1^*=0$. For this commonly used feedforward, a.k.a. input offset or trimming, in LPV gain scheduled control, Condition~$\cvff$ explains why the closed-loop response is stable for the large set-point jump from $\theta_1^*=\pi/4$ to $\theta_1^*=0$, but unstable for the same amount of jump from $\theta_1^*=0$ to $\theta_1^*=-\pi/4$. The instability is mainly due to the violation of Condition~$\cvff$ by the chosen $\kappa^{\mathrm{ff}}$, as an inappropriate design of $\kappa^{\mathrm{ff}}$ can reduce the stability margin obtained from feedback control synthesis. For further analysis of this issue, see Section~\ref{sec:glpv-cmp}. 
    
\subsection{Virtual reference generator} \label{sec:vccm-realization}

As shown in Section~\ref{sec:case-study}, Condition~$\cvff$ is essential to ensure closed-loop stability. However, this condition becomes more restrictive when $\Bc^*$ contains more reference trajectories. Taking the virtual control-affine system \eqref{eq:cfs-diff-dyn} as an example, Condition~$\cvff$ becomes
\begin{equation}\label{eq:cond-ff}
	\dot{x}^*(t)-g(x^*(t),x(t))\in \mathrm{Span}\bigl\{B(x^*(t),x(t))\bigr\},\quad \forall x^*\in \Bc^*, x\in \Bc,
\end{equation}
 where $ \mathrm{Span}\{B\}$ denotes the column space of the matrix $B$. Such a condition often does not hold if the input dimension is smaller than the state dimension.


	
In this section, we relax Condition~$ \cvff $ by introducing virtual reference generators (VRGs). Roughly speaking, instead of forcing $ y^*(t)=x^*(t) $ for all $ t\in\R^+ $, a VRG allows to find a virtual reference $ y^*(t) $ that deviates from $ x^*(t) $, but it represents an admissible virtual state trajectory that converges to $ x^*(t) $, guiding $ x(t) $ towards $ x^*(t) $ as the virtual feedback controller \eqref{eq:virt-fb} ensures that all $y(t)$ (including $x(t)$) converge to $ y^*(t) $.

We choose a copy of the virtual system \eqref{eq:sys-virt} to govern the dynamics of the VRG:
\begin{equation}\label{eq:virt-ref}
	\dot{y}^*=\ff(y^*,x,v^*),
\end{equation}
with initial condition $ y^*(0)=x^*(0) $, where the exogenous signal $ x $ is generated by system \eqref{eq:system}. Let $ d:=x-y^* $, then \eqref{eq:virt-ref} can be rewritten as
\begin{equation} \label{eq:virt-ref-modified}
	\dot{y}^*=F(y^*,v^*,d):=\ff(y^*,y^*+d,v^*).
\end{equation}
Here $ d(t) $ is a known time-varying parameter. As one of our contributions, we replace Condition~{$ \cvff $} with the following weaker condition.
\begin{enumerate}
	\item[$ \cvfv $:] For any reference trajectory $ (x^*,u^*)\in\Bc^* $, there exists a virtual reference controller 
	\begin{equation}\label{eq:VRG-ff-control}
	v^*=\kappa^\mathrm{ff}(y^*,x,x^*,u^*),
	\end{equation}
	with $ u^*=\kappa^\mathrm{ff}(x^*,x^*,x^*,u^*) $, where $ \kappa^\mathrm{ff}:\R^n\times\R^n\times\R^n\times\R^m\rightarrow\R^m $ is a locally Lipschitz continuous map, such that for any signal $ {d}=x-y^* $ with $ |d(t)|\leq Re^{-\lambda t}|d(0)| $, the closed-loop solution $ y^* $ of \eqref{eq:virt-ref-modified} exists and 
	\begin{equation}
	|y^*(t)-x^*(t)|\leq R'e^{-\lambda t}|x(0)-y^*(0)|,
	\end{equation}
	for some constant $ R'>0 $.
\end{enumerate}

\begin{theorem}\label{thm:nonlinear-stabilization}
	Consider the system \eqref{eq:system} and a reference behavior $ \Bc^* $. Assume that there exists a virtual system representation \eqref{eq:sys-virt} such that Condition~$\cvfb$ and $\cvfv$ hold. Then, for any $\kappa^{\mathrm{fb}}$ and $\kappa^{\mathrm{ff}}$ that satisfy $\cvfb$ and $\cvfv$, system \eqref{eq:system} is $ \Bc^* $-universally exponentially stable under the following controller 
	\begin{equation}\label{eq:true-realization-VRG}
	\begin{split}
			\dot{y}^*&=\ff(y^*,x,\kappa^\mathrm{ff}(y^*,x,x^*,u^*)),\quad y^*(0)=x^*(0),\quad
			u=\kappa^\mathrm{fb}(x,x,x^*,\kappa^\mathrm{ff}(y^*,x,x^*,u^*)).
	\end{split}
	\end{equation}
\end{theorem}

\begin{proof}
Condition~{$ \cvfb $} implies $ |x(t)-y^*(t)|\leq Re^{-\lambda t}|x(0)-x^*(0)| $ since $x(t)$ is also a feasible virtual state trajectory of \eqref{eq:sys-virt}. By Condition~{$\cvfv$} we have $ |y^*(t)-x^*(t)|\leq R'e^{-\lambda t}|x(0)-y^*(0)| $, which further leads to $|x(t)-x^*(t)|\leq |x(t)-y^*(t)|+|y^*(t)-x^*(t)|\leq  (R+R')e^{-\lambda t}|x(0)-x^*(0)|$. 
\end{proof}

\begin{example}\label{exmp:cond-1}
		Consider the following nonlinear system
		\begin{equation}\label{eq:sys-exmp-ff}
			\dot{x}=f(x)+Bu:=
			\begin{bmatrix}
				2x_1+x_2 \\ x_1+x_1^2-x_1^3-x_2
			\end{bmatrix}+
			\begin{bmatrix}
				1 \\ 0
			\end{bmatrix}u,
		\end{equation}
		and the reference set $\Bc^*=\{(x^*,u^*)\mid x^*,u^* \text{ are constant signals such that } f(x^*)+Bu^*=0\}$. We construct a control-affine virtual system \eqref{eq:cfs-diff-dyn} with 
		\begin{equation*}
			g(y,x)=
			\begin{bmatrix}
				2y_1+y_2 \\ y_1-y_2+x_1^2-x_1^3
			\end{bmatrix}.
		\end{equation*}
 		Simple incremental analysis yields that the virtual feedback controller $\kappa^{\mathrm{fb}}(y,x,y^*,v^*)=v^*-K(y-y^*)$ with $K=[\ 5\ \ 1.5\ ]$ satisfies Condition~$\cvfb$. For the virtual feedforward design, we can verify that \eqref{eq:cond-ff} does not hold for $g$ since for any set-point $x^*$, there are infinitely many $x_1\in \R $ such that $x_1^*-x_2^*+x_1^2-x_1^3\neq 0$. In fact, Condition $\cvff$ cannot be satisfied for this choice of control-affine virtual embedding, as \eqref{eq:cond-ff} does not hold for any control-affine virtual embedding of \eqref{eq:sys-exmp-ff} that tries to hide the whole or part of the nonlinear term $x_1^2-x_1^3$.  Here we choose the virtual feedforward input $ v^*=u^*=\arg\min_{v\in\R^m}|g(x^*,x)-Bv|^2$. The overall  controller \eqref{eq:true-control} is a linear law:
		\begin{equation}\label{eq:linear}
			u=u^*-K(x-x^*).
		\end{equation} 
		Simulation results in Fig.~\ref{fig:exmp-ff} shows that this controller fails to stabilize the system for some large initial states. Now we consider the following VRG  
		\begin{equation}\label{eq:vrg}
			y_1^*=x_1^*,\quad \dot{y}_2^*=y_1^*-y_2^*+x_1^2-x_1^3,\quad v^*=u^*+x_2^*-y_2^*,
		\end{equation}
		which satisfies Condition~$\cvfv$ as $ y_2^*(t)$ converges to $x_2^*$ if $x_1(t)$ approaches to $y_1^*$. Then, the overall controller \eqref{eq:true-realization-VRG} becomes
		\begin{equation}\label{eq:vrgd}
			\begin{split}
				\dot{y}_2^*=x_1^*-x_2^*+x_1^2-x_1^3, \quad y_2^*(0)=x_2^*,\quad
				u=u^*+x_2^*-y_2^*-K\left(x-\begin{bmatrix}
				x_1^* \\ y_2^*
				\end{bmatrix}\right).
			\end{split}
		\end{equation}
		The closed-loop stability follows by Theorem~\ref{thm:nonlinear-stabilization}, see Fig.~\ref{fig:exmp-ff} for stable response to a large initial state.
		\begin{figure}[!bt]
			\centering
			\includegraphics[width=0.4\linewidth]{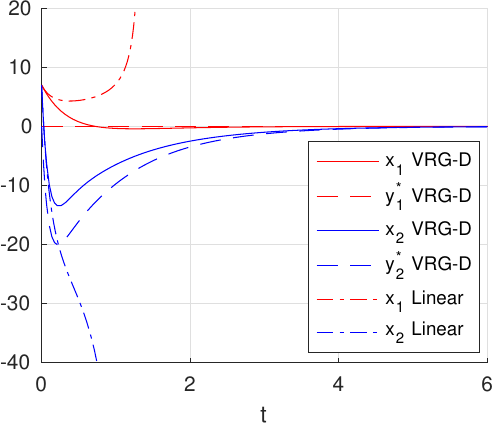}
			\caption{State responses of \eqref{eq:sys-exmp-ff} with VRG-D controller \eqref{eq:vrgd} and linear controller \eqref{eq:linear} from initial state $ x(0)=[\ 7\ \ 7\ ]^\top$. The virtual reference $ y_2^*(t) $ first deviates from $x_2^*$ due to the large initial value of $x_1(t)$ and then converges to $x_2^*$ when $x_1(t)$ approaches to $x_1^*$. The state $x_2(t)$ converges to $x_2^*$ as it follows $ y_2^*(t)$.} \label{fig:exmp-ff}
		\end{figure}
\end{example}

\subsection{Comparison with CCM-based control} \label{sec:CCM:comp}
A special case of the VCCM approach is when we take the original system itself as a virtual representation. In this case, both control synthesis and realization are the same as in the CCM approach \cite{Manchester:2017,Manchester:2018}. With the proposed generalization based on the behavior embedding, our method can be applied to weak stabilization problems where the CCM approach fails to produce a feedback controller, see Example~\ref{exmp:1}. Additionally, the complexity of the control design can be simplified since part of the system nonlinearities can be treated as external parameters at the synthesis stage. 

\begin{example}
	Consider an example taken from \cite{Andrieu2010}, with state $ x=[\ x_1 \ \ x_2 \ \ x_3\ ]^\top $ and nonlinear dynamics $\dot{x}=f(x)+Bu$ where
	\begin{equation}\label{eq:sys-exmp-9}
		f(x)=\begin{bmatrix}
			-x_1+x_3 \\ x_1^2-x_2-2x_1x_3+x_3 \\ -x_2
		\end{bmatrix},\quad B=
		\begin{bmatrix}
			0 \\ 0 \\ 1
		\end{bmatrix}.
	\end{equation}
	The task is set-point tracking, i.e., $\Bc^*=\{(x^*,u^*)\mid f(x^*)+Bu^*=0\}$ with Linear Quadratic Regulation (LQR) cost $\int_0^\infty \bigl(x^\top x+u^2\bigr) dt$. As pointed out in \cite{Manchester:2017}, the above system does not allow for a uniform CCM. Then, a pair of state-dependent controller and metric have been computed by solving state-dependent LMIs. However, implementation of this CCM controller requires solving an online optimization problem at each time step due to the non-uniform metric.  
	
For the proposed VCCM approach, we consider the virtual embedding $ \dot{y}=g(y,x)+Bv$ with 
	\[
		g(y,x)=\begin{bmatrix}
			-y_1+y_3 \\ -y_2+y_3+x_1^2-2x_1x_3 \\ -y_2
		\end{bmatrix}.
	\]
	Since the associated virtual differential dynamics becomes an LTI system, we can obtain a constant metric and an LQR gain $K$ based on the cost functional $\int_0^{\infty}\bigl(\delta_y^\top \delta_y+\delta_v^2\bigr) dt$. We choose the VRG with $y_1^*=x_1^*,\ \dot{y}_2^*=-y_2^*+y_3^*+x_1^2-2x_1x_3,\ y_3^*=x_3^*$ and $ v^*=y_2^*$, which satisfies Condition~$\cvfv$. Then, our method gives a simple dynamic controller for \eqref{eq:sys-exmp-9}:
	\begin{equation}\label{eq:vccm-dyn-ctrl}
		\begin{split}
			\dot{y}_2^*=-y_2^*+x_3^*+x_1^2-2x_1x_3,\quad y_2^*(0)=x_2^*, \quad
			u=y_2^*+K(x-y^*).
		\end{split}
	\end{equation}
	Fig.~\ref{fig:vccm-ccm} shows that for small initial conditions, the responses of VCCM, CCM, and LQR controllers are almost identical, where the LQR controller is constructed based on the local linearized model at the origin. In contrast, for larger initial conditions, the LQR controller was not stabilizing, while both the VCCM and CCM controllers were. On the other hand, the implementation of the VCCM controller \eqref{eq:vccm-dyn-ctrl} is much simpler than the CCM controller in \cite{Manchester:2017} as it does not require online optimization for computing geodesics.
	\begin{figure}[!bt]
		\centering
		\includegraphics[width=0.6\linewidth]{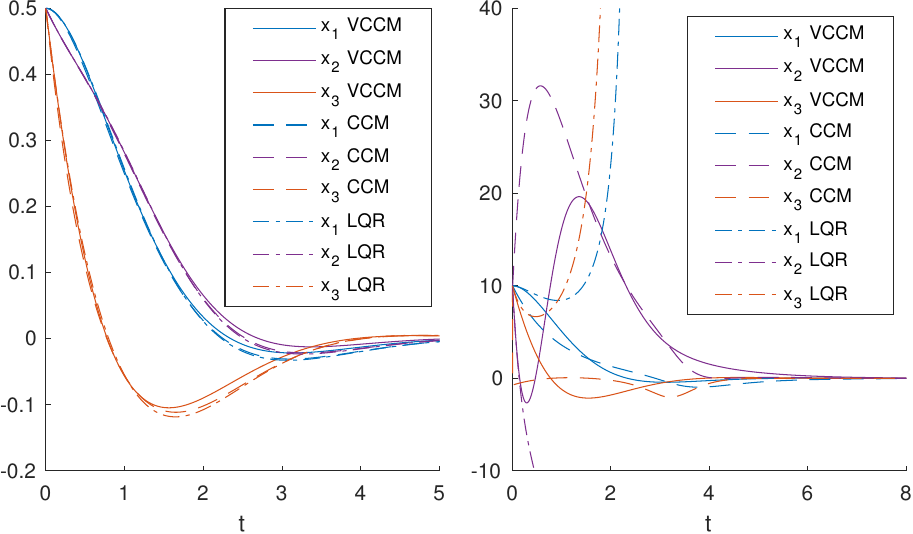}
		\caption{State responses of \eqref{eq:sys-exmp-9} with VCCM, CCM and LQR control from initial state $x(0) = [\ 0.5\ \ 0.5\ \ 0.5\ ]^\top $ (left) and $x(0) = [\ 10\ \ 10\ \ 10\ ]^\top$ (right).}\label{fig:vccm-ccm}
	\end{figure}
	
\end{example}

\section{Disturbance Rejection via Robust VCCM}\label{sec:rejection}

Building on the results of the previous sections, we can extend the VCCM approach for general disturbance rejection. Given the system \eqref{eq:system-pert}, we first construct a virtual system
\begin{equation}\label{eq:sys-virt-pert}
	\dot{y}=\ff(y,x,v,w),\quad \zeta=\hf(y,x,v,w),
\end{equation}
with the property that $ \ff(x,x,u,w)=f(x,u,w) $ and $ \hf(x,x,u,w)=h(x,u,w) $, where $ y(t)\in\R^n,v(t)\in\R^m,\zeta(t)\in\R^q $ are the virtual state, virtual control input and virtual performance output, respectively, and $ x(t) $ is the real state of \eqref{eq:system-pert}. 

The associated differential dynamics of the projected virtual behavior $ \Bv_{x} $ are given by
\begin{equation}\label{eq:diff-sys-pert}
	\begin{split}
		\dot{\delta}_{y} =A(\sigma)\delta_{y}+B(\sigma)\delta_{v}+B_w(\sigma)\delta_{w}, \quad
		\delta_{\zeta} = C(\sigma)\delta_{y}+D(\sigma)\delta_{v}+D_w(\sigma)\delta_{w},
	\end{split}
\end{equation}
where $ \sigma=(y,x,v,w)$, $ A=\frac{\partial \ff}{\partial y} $, $ B=\frac{\partial \ff}{\partial v} $, $ B_w=\frac{\partial \ff}{\partial w} $, $ C=\frac{\partial \hf}{\partial y} $, $ D=\frac{\partial \hf}{\partial v} $ and $ D_w=\frac{\partial \hf}{\partial w} $. Applying the differential state feedback \eqref{eq:diff-controller} to \eqref{eq:diff-sys-pert} gives the closed-loop differential dynamics:
\begin{equation}\label{eq:diff-sys-pert-cl}
	\begin{split}
	\dot{\delta}_{y}=(A+BK)\delta_{y}+B_w\delta_{w}, \quad
	\delta_{\zeta}=(C+DK)\delta_{y}+D_w\delta_{w}.
	\end{split}
\end{equation}
A robust virtual control contraction metric (RVCCM) is a uniformly bounded metric $  M(y,x) $ to establish an $ \Lc_2 $ gain bound for \eqref{eq:diff-sys-pert-cl}, i.e.,
\begin{equation}\label{eq:virt-diff-L2}
	\dot{V}(y,x,\delta_y)\leq -\delta_\zeta^\top \delta_\zeta+\alpha^2\delta_w^\top \delta_w,
\end{equation}
where $ V(y,x,\delta_y)=\delta_y^\top M(y,x)\delta_y $ is called a virtual differential storage function. The following result gives a sufficient condition to search for the pair $ (K,M) $. 
\begin{proposition}
	Suppose that there exist matrix functions $ Y(y,x):\R^n\times\R^n\rightarrow \R^m\times\R^n $ and $ W(y,x):\R^n\times\R^n\rightarrow \R^n\times \R^n $ with $c_2I\succeq W(y,x)\succeq c_1I $ such that
	\begin{equation}\label{eq:rvccm-synthsis}
	\begin{bmatrix}
	H & B_w &(C W+D Y)^\top \\
	B_w^\top  & -\alpha I &   D_w^\top \\
	(CW+DY) &   D_w & -\alpha I
	\end{bmatrix}\prec 0,\quad \forall x,y\in \R^n
	\end{equation}
	where $ H=- \dot{W}+AW+WA^\top+BY+Y^\top B^\top $. Then, the controlled system \eqref{eq:diff-sys-pert-cl} with $ K=YW^{-1} $ satisfies the $ \Lc_2 $-gain condition \eqref{eq:virt-diff-L2} with $ M=W^{-1} $. 
\end{proposition}
\begin{proof}
    The proof follows similarly as in \cite{Manchester:2018}.
\end{proof}
The transformation from \eqref{eq:virt-diff-L2} to \eqref{eq:rvccm-synthsis} is similar to the case of $ H_\infty $ state-feedback control for linear systems (e.g., \cite{Dullerud:2013}). Here \eqref{eq:rvccm-synthsis} takes the form of a point-wise LMI in $ W $ and $ Y $, i.e., it is still convex, but infinite dimensional. Numerically efficient solutions are discussed in Section~\ref{sec:VCCM:concept}.

For realization of the controller in case of disturbance rejection, we consider a VRG of the form
\begin{equation}\label{eq:virt-ref-pert}
	\begin{split}
	\dot{y}^*&=\ff(y^*,x,v^*,w^*),\quad
	\zeta^*=\hf(y^*,x,v^*,w^*),
	\end{split}
\end{equation}
with initial condition $ y^*(0)=x^*(0) $, where the exogenous input $ x $ is generated by the original system \eqref{eq:system-pert} under the control law $ u=\kappa^\mathrm{fb}(x,x,y^*,v^*) $ where $ \kappa^\mathrm{fb} $ is given in \eqref{eq:virt-controller}. From \cite[Th.~1]{Manchester:2018} and the behavioral embedding principle, the real closed-loop system achieves an $ \Lc_2 $-gain bound of $ \alpha $ from $ w-w^* $ to $ z-{\zeta}^* $. We can also obtain the $ \Lc_2 $-gain bound (denoted as $ \alpha_{wy} $) from $ \delta_{w} $ to $ \delta_{y} $ of \eqref{eq:diff-sys-pert-cl} under the differential gain $ K=YW^{-1} $. The performance bound from $ w-w^* $ to $ z-z^* $ is given as follows.

\begin{theorem}\label{thm:performance}
	Consider the system \eqref{eq:system-pert} and a reference behavior $\Bc^*$. Assume that there exists a virtual embedding such that \eqref{eq:rvccm-synthsis} is feasible and  there exists a virtual feed-forward controller $ \kappa^\mathrm{ff} $ achieving $ \Lc_2 $ gain of $ \alpha_{y\zeta} $ from $ x-x^* $ to $ {\zeta}^*-z^* $ for the VRG \eqref{eq:virt-ref-pert}. Then, the realization of the overall control law \eqref{eq:true-realization-VRG} achieves a $ \Bc^* $-universal $ \Lc_2 $-gain bound of $ \widetilde{\alpha}=\sqrt{\alpha^2+\alpha_{wy}^2\alpha_{y\zeta}^2} $ for the system \eqref{eq:system-pert}.
\end{theorem} 

\begin{proof}
	From the above analysis, the upper bound of the $\Lc_2$ gain can be established by
	\begin{equation*}
	\begin{split}
	\|(z-z^*)_T\|_2^2&\leq \|(z-{\zeta}^*)_T\|_2^2+\|({\zeta}^*-z^*)_T\|_2^2 \leq \alpha^2\|(w-w^*)_T\|_2^2+\alpha_{y\zeta}^2\|(x-x^*)_T\|_2^2 \\
	&\leq (\alpha^2+\alpha_{wy}^2\alpha_{y\zeta}^2)\|(w-w^*)_T\|_2^2. 
	\end{split}
	\end{equation*}
\end{proof}

\section{A unified generalization for LPV control}\label{sec:comparison}

In this section we first summarize the design procedure of the proposed VCCM approach. Then, by step-by-step comparisons we show that the VCCM approach is a unified generalization for two distinct categories of LPV control methods, namely global and local LPV state-feedback control. Moreover, through the lens of Theorem~\ref{thm:vccm} we provide explanations to the question why both the local and global standard LPV control realizations do not provided rigorous global stability and performance guarantees for tracking control.

\subsection{Overview of VCCM-based control}

\begin{figure}[!bt]
	\centering\includegraphics[width=0.7\linewidth]{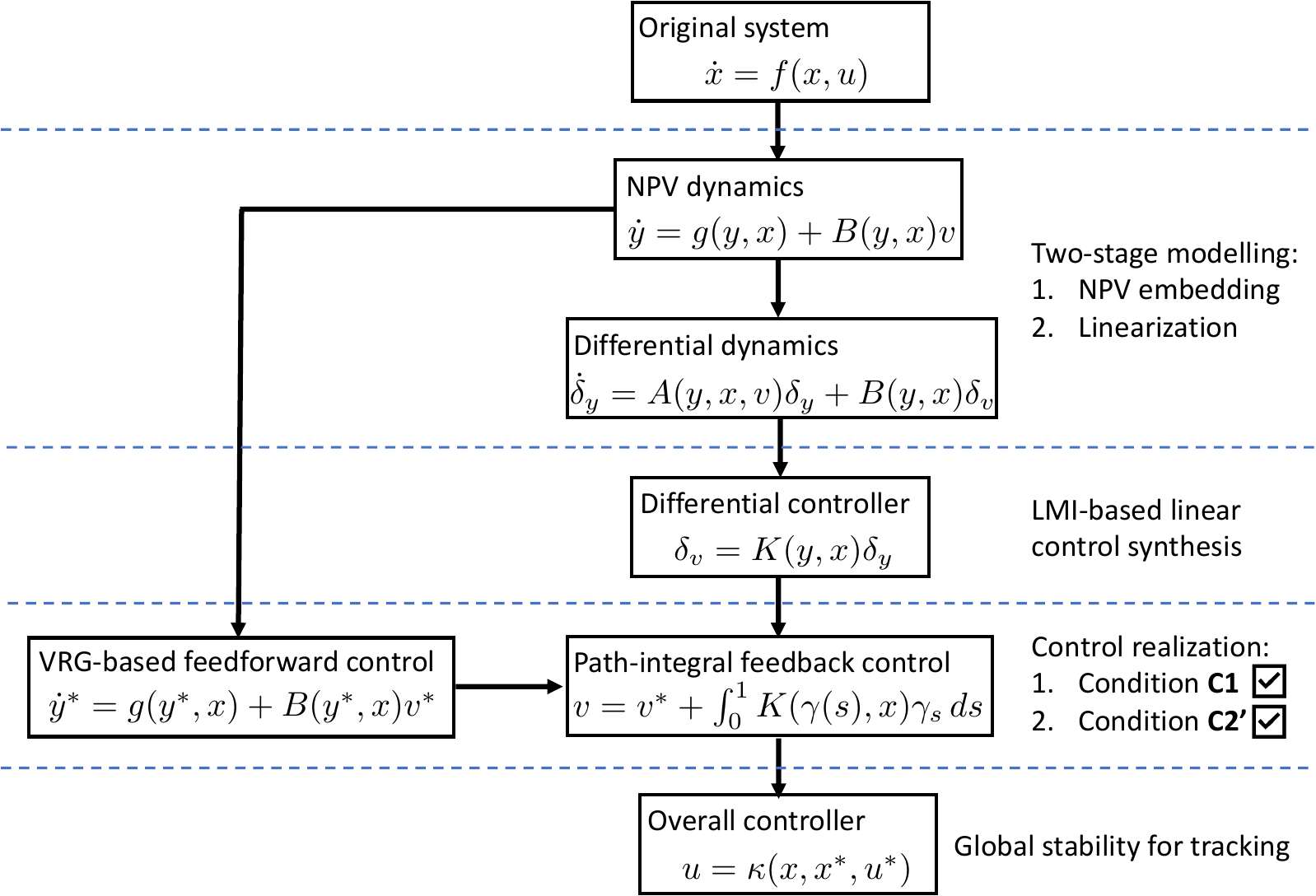}
	\caption{Control design procedure for the proposed VCCM approach.}\label{fig:vccm}
\end{figure}

We summarize the procedure of the VCCM-based control as follows (see also Fig.~\ref{fig:vccm}):
\begin{itemize}
    \item \emph{Two-stage modelling}: We first choose a virtual system by treating some state-dependent terms as exogenous parameters, resulting a nonlinear parameter-varying (NPV) behavior embedding. Then, by taking continuous linearization w.r.t. the virtual state and input, we obtain a virtual differential dynamics, which can be treated as an LPV system.
    \item \emph{Control synthesis}: We use the LMI-based control synthesis framework for LPV systems to design a linear differential state-feedback controller.
    \item \emph{Control realization}: We construct a virtual feedback controller by integrating the differential controller over a particular path. This realization ensures that Condition~$\cvfb$ holds. We also need to design a virtual feedforward controller based on the NPV embedding such that either Condition~$\cvff$ or $\cvff'$ holds. Finally, the overall controller based on behavior embedding principle provides rigorous stability and performance guarantees. 
\end{itemize}


\subsection{Local LPV control}

We rephrase a classic local LPV control method -- the equilibrium linearization approach \cite{Shamma90c} in the VCCM framework. For simplicity, we only consider control design for the nominal system \eqref{eq:system}.

\subsubsection{Two-stage modelling.} 
\begin{figure}[!bt]
	\centering
	\begin{tabular}{ccc}
		\includegraphics[width=0.3\linewidth]{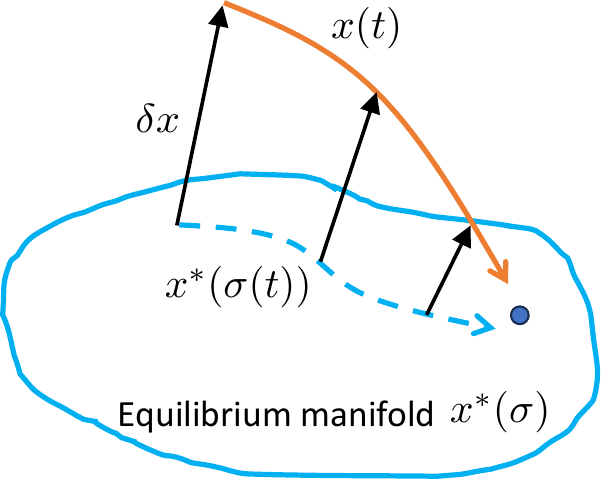} & $\quad$&
	\includegraphics[width=0.3\linewidth]{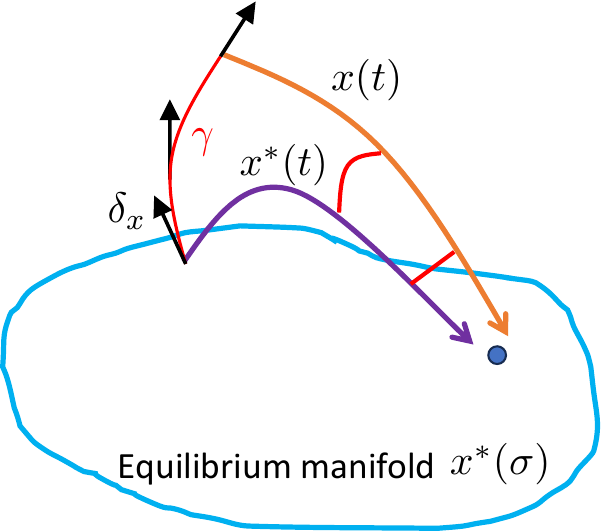} \\
	{\scriptsize (a) Gain scheduling} & $\quad$& {\scriptsize (b) VCCM}
	\end{tabular}
	\caption{Geometric illustration of different control realizations.}\label{fig:realization}
\end{figure}

The local LPV control method takes a \emph{trivial} NPV embedding, i.e., the virtual dynamics is a copy of the original nonlinear dynamics \eqref{eq:system}.
Let $ \{(x^*,u^*)(\sigma) \}_{\sigma\in\Omega} $ be a smooth equilibrium family  of \eqref{eq:system} where $ \sigma \in \Omega \subset \R^{n_\sigma}$ is a scheduling variable depending on the internal state $x$, i.e., $\sigma=\psi(x)$ is a smooth mapping. By performing Jacobian linearization of \eqref{eq:system} around the equilibrium family, we obtain an LPV model as follows:
\begin{equation}\label{eq:nominal-sys2}
\frac{d}{dt} \delta x= A(\sigma) \delta x + B(\sigma)\delta u:=\frac{\partial f}{\partial x}(x^*(\sigma), u^*(\sigma))\delta x+\frac{\partial f}{\partial u}(x^*(\sigma), u^*(\sigma)) \delta u,\; \sigma\in\Omega
\end{equation}
where $\delta x=x-x^*(\sigma)$,  $\delta u=u-u^*(\sigma)$ are the \emph{deviation variables}. The VCCM approach extends the modelling of local LPV control in two aspects. First, it allows for more general NPV embeddings, which can hide some ``trouble'' nonlinear terms into the exogenous parameters and result in a simple control synthesis problem. Secondly, linearization is extended from the equilibrium manifold to the entire state/input space (see Fig~\ref{fig:realization}), which allows for tracking control design for time-varying references.

\subsubsection{Control synthesis.} By solving the same control synthesis problem \eqref{eq:vccm-synthesis} as the VCCM approach, we obtain an LPV controller of the form
\begin{equation}\label{eq:LPV-control}
\delta u=K(\sigma)\delta x,
\end{equation}
with a given parameterization of $K$ (e.g. an affine function of $\sigma$). That is, there exists an Lyapunov function $V(\delta x)=\delta x^\top M(\sigma) \delta x$ with $M(\sigma)\succ 0$ such that 
\begin{equation}\label{eq:Lyapunov-local-lpv}
    \frac{d}{dt}V(\delta x)=\delta x^\top \overset{P(\sigma)}{\overbrace{\left[\frac{\partial M}{\partial \sigma} \dot{\sigma}+A_c(\sigma)^\top M(\sigma)+ M(\sigma) A_c(\sigma)\right]}} \delta x < 0
\end{equation}
for all $\delta x\neq 0$, where $A_c(\sigma)=A(\sigma)+B(\sigma)K(\sigma)$.

\subsubsection{Control realization.}
The realization task for a local LPV controller \eqref{eq:LPV-control} is to construct a gain-scheduled law $ u=\kappa(x,\sigma) $ for \eqref{eq:system} satisfying 
\begin{subequations}
    \begin{gather}
    u^*(\sigma)=\kappa(x^*(\sigma),\sigma), \label{eq:GS-cond-1}\\
        \frac{\partial \kappa}{\partial y}(y^*(\sigma),\sigma)=K(\sigma). \label{eq:GS-cond-2}
    \end{gather}
\end{subequations}
Condition \eqref{eq:GS-cond-1} means that the reference point $x^*$ is also an equilibrium of the closed-loop system. Condition \eqref{eq:GS-cond-2} states that linearization of $ u=\kappa(x,\sigma) $ at this equilibrium is the local LPV controller \eqref{eq:LPV-control}, which is the key to guarantee closed-loop stability. However, such a realization generally does not exist when $K(\psi(x))$ is not completely-integrable. A typical realization for \eqref{eq:LPV-control} used in the LPV literature is 
\begin{equation} \label{eq:GS-control}
u=u^*(\sigma)+K(\sigma)\bigl(x-x^*(\sigma)\bigr),
\end{equation}
leading to a nonlinear controller of the form
\begin{equation}\label{eq:GS-control-implementation}
u=u^*(\psi(x))+K(\psi(x))\bigl(x-x^*(\psi(x))\bigr).
\end{equation}

As pointed out in \cite{Rugh:2000}, the closed-loop stability of the above gain-scheduling controller suffers from two drawbacks: First, the state $x$ needs to be sufficiently close to the equilibrium manifold $x^*(\sigma)$; Second, the scheduling signal $\sigma$ needs to be sufficiently ``slowly varying''. We now give explanations to the above issues by checking the conditions of Theorem~\ref{thm:vccm}. 

Firstly, we show that Condition~$\cvfb$ may not hold for the feedback controller \eqref{eq:GS-control-implementation}.  Although $ \sigma $ is implicitly involved in \eqref{eq:GS-control-implementation} via equilibrium parameterization, linearization of \eqref{eq:GS-control-implementation} at the equilibrium manifold yields
\begin{equation*}
\begin{split}
\delta u\!=\!K(\sigma)\delta x 
+\!\overset{K_h(\sigma)}{\overbrace{\left(\frac{\partial u^*(\sigma)}{\partial\sigma}\!-\!K(\sigma)\frac{\partial x^*(\sigma)}{\partial \sigma}\right)\!\frac{\partial \psi}{\partial x}(x^*(\sigma))}}\delta x.
\end{split}
\end{equation*}
Compared with \eqref{eq:LPV-control}, the above equation contains a so-called \emph{hidden coupling term} $ K_h(\sigma)\delta x$. Then, the derivative of Lyapunov function in \eqref{eq:Lyapunov-local-lpv} becomes 
\begin{equation}
    \frac{d}{dt}V(\delta x)=\delta x^\top P(\sigma)\delta x+\delta x^\top \left[B(\sigma)K_h(\sigma)^\top M(\sigma)+M(\sigma)B(\sigma)K_h(\sigma)\right]\delta x,
\end{equation}
which may not always be negative. Thus, Condition~$\cvfb$ may be violated by the virtual feedback realization \eqref{eq:GS-control-implementation}, regardless the fact that exponential stability is achieved in the local control synthesis step. This could lead to performance deterioration or even closed-loop instability  \cite{Rugh:2000}. 

Then, we explain the necessity of sufficiently ``slow-varying" scheduling signals  for closed-loop stability through the lens of Condition~$\cvff$. The feedforward component in \eqref{eq:GS-control-implementation} is $x^*(t)=x^*(\sigma(t))$ and $u^*(t)=u^*(\sigma(t))$. Substituting $ (x^*,u^*)(\sigma(t)) $ into \eqref{eq:system} yields a residual term $ \frac{\partial x^*(\sigma)}{\partial \sigma} \dot{\sigma} $, i.e., Condition~$\cvfb$ does not hold as $ (x^*,u^*)(\sigma(t)) $ is not an admissible trajectory to the virtual system \eqref{eq:system}. If $\dot{\sigma}$ is not sufficiently small, the residual term can drive the state away from the close neighborhood of $ x^*(\sigma) $, where the local stability guarantees of the design hold.

Compared to gain-scheduling, the proposed VCCM control scheme achieves rigorous global stability and performance guarantees based on the following two facts. Firstly, it does not contain any residual term as its reference trajectory is a feasible solution to the system dynamics. Secondly, thanks to continuous linearization and contraction theory, it does not suffer from hidden coupling effects as it only requires $K$ to be path-integrable. By integrating the differential controller along geodesics, it achieves the designed stability and performance \cite{Manchester:2017}. 

We argue that the proposed path-integral based realization is the correct realization of gain scheduling that has been searched for in the past. Considering a linear differential controller $\delta_u=K(x)\delta_x$, its path-integral realization is $v=\nu(1)$ where $\nu:[0,1]\rightarrow\R^m$ is a control path satisfying
\begin{equation}\label{eq:path-integral}
	\nu(s):=v^*+\int_{0}^{s}K(\gamma(\tau))\gamma_s(\tau)d\tau
\end{equation} 
with $\gamma:[0,1]\rightarrow\R^n$ as a geodesic connecting $y^*$ to $y$. Note that Condition \eqref{eq:GS-cond-1} holds as $ \nu(1)=\nu(0)=u^*$ if $x=x^*$. Condition \eqref{eq:GS-cond-2} holds along the path $\gamma$, i.e., $ \nu_s=K(\gamma(s))\gamma_s(s)$ for all $s\in [0,1]$. Then, we connect it to the gain-scheduling controller \eqref{eq:GS-control} by taking $N$ sampled points of $\gamma$ where $N$ is sufficiently large. A simple numerical integration scheme for the path integral \eqref{eq:path-integral} is
\begin{equation*}
\nu(s_{j+1})\approx \nu(s_{j})+K(\gamma(s_j))\bigl( \gamma(s_{j+1})-\gamma (s_j)\bigr),
\end{equation*}
where $v(s_0)=v^*$. Based on this observation, the control path $ \nu $ integrates a series of gain-scheduling controllers \eqref{eq:GS-control} along the path $ \gamma $, which corresponds to a ``scheduled path'' to reach reference trajectory, and the VCCM control realization is the corresponding control action to the state  $ x=\gamma(1) $ at one end-point of the path. 


Here we give a comparison example of the gain-scheduling and path-integral realization schemes.
\begin{example}
	Consider the following nonlinear system from \cite{Rugh:1991}:
\begin{equation}\label{eq:system-example-1}
\dot{x}_1=-x_1-x_2+r,\quad \dot{x}_2=1-e^{-x_2}+u,
\end{equation}
where $ r(t) $ is a measurable reference. Define the equilibrium family by $ x^*(r)=(0,r) $ and $ u^*(r)=e^{-r}-1 $, and introduce the scheduling variable $\sigma=e^{-r}$. Pole placement at $ \lambda_{1,2}=-2 $ gives an LPV controller \eqref{eq:LPV-control} with $ K(\sigma)={[\ 1\ \ (-3-\sigma)\ ]}$. Substituting $\sigma=e^{-r}$ into \eqref{eq:GS-control} gives a gain-scheduled controller (denoted by GSC 1):
\begin{equation}\label{eq:lpv-controller-1}
u=e^{-r}-1+x_1-(3+e^{-r})(x_2-r).
\end{equation}
Unstable closed-loop responses are observed in Fig.~\ref{fig:cmp}. Alternatively, applying the equilibrium relation $\sigma=e^{-x_{2}}$ gives a controller (denoted by GSC 2):
\begin{equation}\label{eq:comp-sche}
	u=x_1+e^{-x_2}-1,
\end{equation}
which contains a hidden coupling term with $K_h=[\ 0\ \ 3\ ]^\top$. As a direct consequence, the closed-loop differential dynamics have eigenvalues with larger real parts, i.e., $ \lambda_{1,2}=-1/2\pm\sqrt{3}/2i $, leading to slower responses as shown Fig.~\ref{fig:cmp}. Moreover, due to the non-zero residual term, the above controller cannot achieve error-free tracking for time-varying references $r(t)$. The proposed realization \eqref{eq:virt-controller} takes a path integral of the local LPV controller along a geodesic $ \gamma $ (in this case, $ \gamma $ is a straight line connecting $x^*=[0,r]^\top$ to $x$), which can be analytically computed via
\begin{equation}\label{eq:contraction-control-1}
\begin{split}
u=u^*+\int_{0}^{1}K(\gamma(s))(x-x^*)ds=\dot{r}+x_1-3(x_2-r)+e^{-x_2}-1,
\end{split}
\end{equation}
where $ u^*=e^{-r}-1+\dot{r} $. Fig.~\ref{fig:cmp} shows that it can follow both piecewise set-point and time-varying reference.  

\begin{figure}[!bt]
	\centering\includegraphics[width=0.5\linewidth]{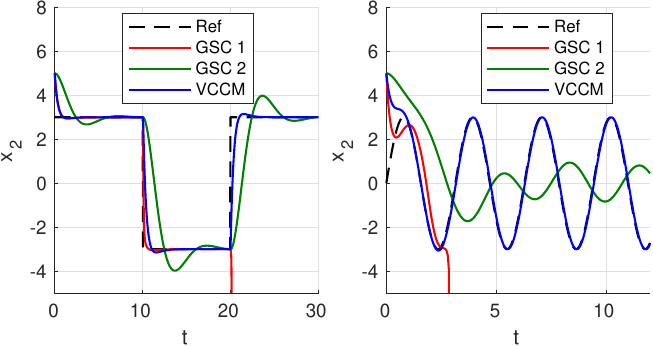}
	\caption{State responses of \eqref{eq:system-example-1} with gain-scheduling and VCCM control for set-point and reference tracking.}\label{fig:cmp}
\end{figure}
\end{example}

Among other local LPV control methods, the so-called \emph{velocity linearization} approach \cite{Leith98} uses a similar technique as continuous linearization, i.e., it uses time derivatives of $u(t)$ and $x(t)$ to produce $\delta_u$ and $\delta_x$, respectively. Then, the LPV control design is executed on this form and the controller is intuitively, realized by integrating its output $\delta_{u}$ and fed by differentiating the state to produce $\delta_{x}$. However, the controlled system can converge to an arbitrary response and suffers from implementation issues, due to required differentiation of the measured/estimated state signal. The work \cite{Mehendale2006} proposed a gain scheduling law based on the velocity form as a solution to the realization problems, but without any theoretical analysis of its implications on stability and performance of the resulting closed loop system. Moreover, this approach will suffer from convergence problems as it fails to guarantee Condition~$\cvff$. 

\subsection{Global LPV control} \label{sec:glpv-cmp}
Similarly, we rephrase the global LPV control approach \cite{shamma1991guaranteed} in the VCCM framework. 
\subsubsection{Two-stage modelling.} In global LPV control, the applied LPV embedding principle can be understood in our setting as follows: the nonlinear system model \eqref{eq:system} is rewritten as a virtual system 
\begin{equation}\label{eq:sys-LPV}
\dot{y}=\ff(y,x,v):=\hat A(x)y + \hat B(x)v, 
\end{equation}
with $ \ff(x,x,u)=f(x,u) $. The main difference w.r.t. the NPV embedding \eqref{eq:sys-virt} is that \eqref{eq:sys-LPV} has to be linear in $(y,v)$. Note that construction of \eqref{eq:sys-LPV} is not unique and in some cases existence of such a virtual form requires dependence of $\hat A$ and $\hat B$ on $u$ as well. 

By introducing a scheduling map $\sigma=\psi(x)$, where $\psi$ is a vector function, such that $\hat{A}(x)=A(\sigma)$ and $\hat{B}(x)=B(\sigma)$ are affine, polynomial or rational functions of $\sigma$, and restricting the variation of $\sigma(t)$ to a convex set $\Omega\subset \R^{n_\sigma}$, a global LPV model of \eqref{eq:system-pert} is formulated as
\begin{equation}\label{eq:nominal-sys3}
\dot y = A(\sigma) y + B(\sigma)u,\quad \sigma\in\Omega.
\end{equation}
From the modeling viewpoint, the VCCM approach extends the LPV embedding to NPV embeddings, which can reduce the conservativeness of the behavioral embedding principle. Recent research results in the LPV literature also try to formulate extensions of the available toolset in this direction \cite{Rotondoi:2019,Sala:2019}.

\subsubsection{Control synthesis.} Based on the same synthesis formulation \eqref{eq:vccm-synthesis} of the VCCM approach, we can obtain an LPV controller 
\begin{equation}\label{eq:LPV-control2}
v=K(\sigma)y,
\end{equation}
with a given parameterization of $K$ (e.g. an affine function of $\sigma$).

\subsubsection{Control realization.} Due to the behavior embedding principle, the real controller takes the form of
\begin{equation}
	u=K(\psi(x))x,
\end{equation}
which stabilizes \eqref{eq:nominal-sys3} if it maintains $\sigma(t)\in \Omega$. For set-point tracking, a possible realization (\cite{Scorletti:2015,Koelewijn:2020}) of \eqref{eq:LPV-control2} is 
\begin{equation}\label{eq:LPV-control3}
	u=u^*+K(\psi(x))(x-x^*).
\end{equation} 
It is assumed that due to the linearity of \eqref{eq:nominal-sys3}, stability guarantee for the origin trivially extends to any other set-point $(x^*,u^*)$ by applying a state transformation in the constructed Lyapunov / storage function $V$. However, it has been recently shown that there is a fundamental gap in the reasoning extending the resulting Lyapunov / storage function for such cases \cite{Koelewijn:2020}. 

For the LPV embedding \eqref{eq:sys-LPV}, the VCCM synthesis does not require any Jacobian linearization, and hence it is equivalent to the LPV synthesis. Since the metric is independent of $y$, the realization for \eqref{eq:LPV-control2} takes the form of
\begin{equation}\label{eq:vccm-lpv-control}
	u=\kappa^{\mathrm{ff}}(x,x^*,u^*)+K(\psi(x))(x-x^*),
\end{equation}
where the feed-forward term $\kappa^{\mathrm{ff}}$ is chosen such that Condition~$\cvff$ holds. From Theorem~\ref{thm:vccm} and \ref{thm:performance}, the above controller achieves global stability and performance for all set-points. Thus, the possible loss of stability and performance guarantees of the LPV realization \eqref{eq:LPV-control3} is due to the fact that Condition~$\cvff$ generally does not hold for the virtual feed-forward controller $v^*=u^*$. Here we give a brief explanation, see \cite{Koelewijn:2020} for details. 

For simplicity, we consider a simple LPV embedding:
\begin{equation}
	\dot{y}=A(\sigma)y+Bv,\quad \sigma=\psi(x).
\end{equation}
Suppose that there exist an LPV controller $ v=K(\sigma)y $ and a quadratic Lyapunov function $ V(y)=y^\top M y $ with $ M\succ 0 $ such that the closed-loop system $ \dot{y}=A_\mathrm{c}(\sigma)y:=\bigl(A(\sigma)+BK(\sigma) \bigr)y $ is asymptotically stable, i.e., $\forall \sigma\in\Omega$
\begin{equation}
	P(\sigma):=A_\mathrm{c}^\top\! (\sigma) M+M A_\mathrm{c}(\sigma)\prec 0.
\end{equation} 
For the set-point $ (x^*,u^*)\neq (0,0) $, the closed-loop system under the realization \eqref{eq:LPV-control3} becomes
\begin{equation}
    \dot{x}=A(x)x+B[u^*+K(\psi(x))(x-x^*)].
\end{equation}
By introducing $ \delta x=x-x^* $, the closed-loop error dynamics can be rewritten as
\begin{equation}\label{eq:LPV-cl}
	\begin{split}
	\dot{\delta x}&=A(\sigma)x-A(\sigma^*)x^*+BK(\sigma)\delta x=\bigl(A_\mathrm{c}(\sigma)+\Delta(x,x^*)\bigr)\delta x
	\end{split}
\end{equation}
where $ \Delta $ is defined by $ \Delta(x,x^*)(x-x^*):=\bigl(A(\sigma)-A(\sigma^*)\bigr)x^* $ with $ \sigma^*=\psi(x^*) $. By using the quadratic Lyapunov function $ V(\delta x)=\delta x^\top M\delta x $, we obtain
\begin{equation}
	\dot{V}(\delta x)=\delta x^\top\left(P+\Delta^{\!\top}\! M+M\Delta\right)\delta x.
\end{equation}
If there exist $ (x,x^*) $ with $ \psi(x),\psi(x^*)\in\Omega $ such that $ P_\Delta(x,x^*):=P(\psi(x))+\Delta^\top(x,x^*) M+M\Delta(x,x^*) $ is indefinite, then the stability guarantee of the origin cannot be extended to the set-point $ (x^*,u^*) $. Performance deterioration or even instability can be observed when $ \Delta $ is sufficiently large, see Example~\ref{exmp:glpv-vccm}. However, for the VCCM-based realization \eqref{eq:vccm-lpv-control}, the term $ \Delta(x,x^*) $ in \eqref{eq:LPV-cl} can be compensated by the feed-forward law $ \kappa^\mathrm{ff} $ since Condition~{$ \cvff $} implies 
\begin{equation}
	A(\psi(x))x^*+B\kappa^\mathrm{ff}(x,x^*,u^*)=0.
\end{equation}
Thus, quadratic stability guarantee of the origin is correctly extended to the set-point $ (x^*,u^*) $. 

Examples have been reported in \cite{Scorletti:2015,Koelewijn:2019} to show the loss of $ \Lc_2 $-gain performance guarantees for set-point tracking when an external disturbance is present. Here, we give an example where, besides loss of performance guarantees, closed-loop instability occurs.

\begin{example}\label{exmp:glpv-vccm}
	Consider the nonlinear scalar system:
	\begin{equation}\label{eq:scalar-sys}
		\dot{x}=f(x)+u:=-x+x^3+u,
	\end{equation}
	and its LPV embedding $\dot{y}=-(1-\sigma)y+v$ where $ \sigma=x^2 $. We obtain an LPV control gain $ K(\sigma)=-2.49-1.01\sigma$ by solving \eqref{eq:rvccm-synthsis} with $ \alpha=0.5 $, $ B_w=0 $, $ D_w=0 $, $ C=[\ 1\ \ 0\ ]^\top $, $ D=[\ 0\ \ 0.1\ ]^\top $ and $ \Omega=[0, 4] $. The closed-loop responses of the global LPV (GLPV) controller \eqref{eq:LPV-control3} and VCCM-based controller \eqref{eq:vccm-lpv-control} for the reference behavior $\Bc^*=\{(x^*,u^*)\mid u^*=f(x^*)\}$ are depicted in Fig.~\ref{fig:response}. When the set-point $x^*$ is the origin, both controllers yield the same closed-loop response. When the magnitude of $x^*$ increases, undesired responses (e.g., convergence to a different equilibrium or instability) can be observed with the global LPV controller. 
	
\begin{figure}[!bt]
		\centering
		\includegraphics[width=0.6\linewidth]{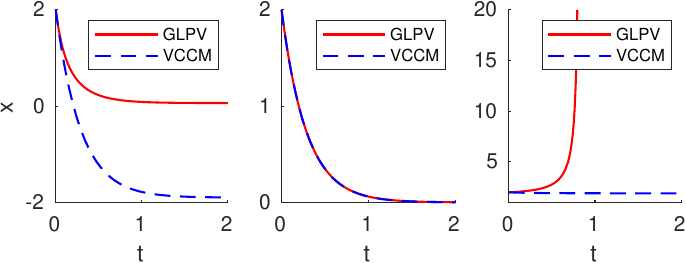}
		\caption{State responses of \eqref{eq:scalar-sys} with GLPV and VCCM control for different set-points: $x^*=-1.9$ (left), $x^*=0$ (middle), and $x^*=1.9$ (right). For the special case with $x^*=0$, the VCCM controller is identical as the GLPV controller. For other reference points, the GLPV control may converge to a different equilibrium or yield unstable closed-loop behavior while our VCCM controller converges to the desired equilibrium. }\label{fig:response}
	\end{figure}	
\end{example}

\section{Conclusion}
In this paper, we present a nonlinear state feedback design approach based on the concept of \emph{virtual control contraction metrics} (VCCMs). This approach can be understood as the generalization of several nonlinear control methods. Firstly, it has weaker assumptions than the existing virtual contraction based control methods. Secondly, it generalizes the CCM based control design to weak stabilization problems. Finally, it is also a unified generalization of two distinct categories of LPV state-feedback control methods. The VCCM control realization provides rigorous stability and performance guarantees for tracking problems while such properties generally do not hold for conventional LPV-based control realizations.

\bibliographystyle{ieeetr}
\bibliography{vccm-ref}

\begin{thebibliography}{10}

\bibitem{Sontag:1983}
E.~Sontag, ``A {L}yapunov-like characterization of asymptotic controllability,'' {\em SIAM Journal on Control and Optimization}, vol.~21, no.~3, pp.~462--471, 1983.

\bibitem{Artstein:1983}
Z.~Artstein, ``Stabilization with relaxed controls,'' {\em Nonlinear Analysis: Theory, Methods \& Applications}, vol.~7, no.~11, pp.~1163--1173, 1983.

\bibitem{Sontag:1989}
E.~Sontag, ``A `universal' construction of {A}rtstein's theorem on nonlinear stabilization,'' {\em Systems \& control letters}, vol.~13, no.~2, pp.~117--123, 1989.

\bibitem{Slotine:1991}
J.-J. Slotine and W.~Li, {\em Applied nonlinear control}, vol.~199.
\newblock Prentice hall Englewood Cliffs, NJ, 1991.

\bibitem{Krstic:1995}
M.~Krstic, I.~Kanellakopoulos, and P.~Kokotovic, {\em Nonlinear and adaptive control design}.
\newblock Wiley New York, 1995.

\bibitem{Boyd:1994}
S.~Boyd, L.~El~Ghaoui, E.~Feron, and V.~Balakrishnan, {\em Linear matrix inequalities in system and control theory}, vol.~15.
\newblock SIAM, 1994.

\bibitem{Rantzer:2001}
A.~Rantzer, ``A dual to {L}yapunov's stability theorem,'' {\em Systems \& Control Letters}, vol.~42, no.~3, pp.~161--168, 2001.

\bibitem{Lasserre:2008}
J.~B. Lasserre, D.~Henrion, C.~Prieur, and E.~Tr{\'e}lat, ``Nonlinear optimal control via occupation measures and lmi-relaxations,'' {\em SIAM Journal on Control and Optimization}, vol.~47, no.~4, pp.~1643--1666, 2008.

\bibitem{Manchester:2017}
I.~Manchester and J.-J. Slotine, ``Control contraction metrics: Convex and intrinsic criteria for nonlinear feedback design,'' {\em IEEE Transaction Automatic Control}, vol.~62, no.~6, pp.~3046--3053, 2017.

\bibitem{Manchester:2018}
I.~Manchester and J.-J. Slotine, ``Robust control contraction metrics: {A} convex approach to nonlinear state-feedback ${H}_\infty$ control,'' {\em IEEE Control System Letter}, vol.~2, no.~3, pp.~333--338, 2018.

\bibitem{Lohmiller:1998}
W.~Lohmiller and J.-J. Slotine, ``On contraction analysis for non-linear systems,'' {\em Automatica}, vol.~34, pp.~683--696, 1998.

\bibitem{Hines:2011}
G.~Hines, M.~Arcak, and A.~Packard, ``Equilibrium-independent passivity: {A} new definition and numerical certification,'' {\em Automatica}, vol.~47, pp.~1949--1956, 2011.

\bibitem{Simpson-Porco:2019}
J.~Simpson-Porco, ``Equilibrium-independent dissipativity with quadratic supply rates,'' {\em IEEE Transactions Automate Control}, vol.~64, no.~4, pp.~1440--1455, 2019.

\bibitem{Rugh:2000}
W.~Rugh and J.~Shamma, ``Research on gain scheduling,'' {\em Automatica}, vol.~36, pp.~1401--1425, 2000.

\bibitem{Hoffmann:2015}
C.~Hoffmann and H.~Werner, ``A survey of linear parameter-varying control applications validated by experiments or high-fidelity simulations,'' {\em IEEE Transactions on Control System Technology}, vol.~23, no.~2, pp.~416--433, 2015.

\bibitem{Rugh:1991}
W.~J. Rugh, ``Analytical framework for gain scheduling,'' {\em IEEE Control Systems Magazine}, vol.~11, pp.~74--84, 1991.

\bibitem{Fromion:2003}
V.~Fromion and G.~Scorletti, ``A theoretical framework for gain scheduling,'' {\em International Journal Robust Nonlinear Control}, vol.~13, no.~10, pp.~951--982, 2003.

\bibitem{Scorletti:2015}
G.~Scorletti, V.~Fromion, and S.~De~Hillerin, ``Toward nonlinear tracking and rejection using lpv control,'' {\em IFAC-PapersOnLine}, vol.~48, no.~26, pp.~13--18, 2015.

\bibitem{Koelewijn:2020}
P.~Koelewijn, G.~S. Mazzoccante, R.~T{\'o}th, and S.~Weiland, ``Pitfalls of guaranteeing asymptotic stability in lpv control of nonlinear systems,'' in {\em 2020 European Control Conference (ECC)}, pp.~1573--1578, IEEE, 2020.

\bibitem{Jouffroy:2004}
J.~Jouffroy and J.-J. Slotine, ``Methodological remarks on contraction theory,'' in {\em 2004 43rd IEEE Conference on Decision and Control (CDC)(IEEE Cat. No. 04CH37601)}, vol.~3, pp.~2537--2543, IEEE, 2004.

\bibitem{Wang:2005}
W.~Wang and J.-J. Slotine, ``On partial contraction analysis for coupled nonlinear oscillators,'' {\em Biological cybernetics}, vol.~92, no.~1, pp.~38--53, 2005.

\bibitem{Manchester:2018a}
I.~Manchester, J.~Tang, and J.-J. Slotine, ``Unifying robot trajectory tracking with control contraction metrics,'' in {\em Robotics Research}, pp.~403--418, Springer, 2018.

\bibitem{Reyes:2020}
R.~Reyes-B\'{a}ez, A.~van~der Schaft, B.~Jayawardhana, and L.~Pan, ``A family of virtual contraction based controllers for tracking of flexible-joints port-hamiltonian robots: Theory and experiments,'' {\em International Journal of Robust and Nonlinear Control}, vol.~30, no.~8, pp.~3269--3295, 2020.

\bibitem{wang:2021}
R.~Wang, P.~J.~W. Koelewijn, R.~T{\'o}th, and I.~R. Manchester, ``Nonlinear parameter-varying state-feedback design for a gyroscope using virtual control contraction metrics,'' {\em International Journal of Robust and Nonlinear Control}, vol.~31, no.~17, pp.~8147--8164, 2021.

\bibitem{Wang:2017}
R.~Wang, I.~Manchester, and J.~Bao, ``Distributed economic mpc with separable control contraction metrics,'' {\em IEEE Control Syst. Lett.}, vol.~1, pp.~104--109, 2017.

\bibitem{Forni:2014}
F.~Forni and R.~Sepulchre, ``A differential {L}yapunov framework for contraction analysis,'' {\em IEEE Transactions Automatic Control}, vol.~3, no.~59, pp.~614--628, 2014.

\bibitem{Aminzare:2014}
Z.~Aminzare and E.~D. Sontagy, ``Contraction methods for nonlinear systems: A brief introduction and some open problems,'' in {\em 53rd IEEE Conference on Decision and Control}, pp.~3835--3847, IEEE, 2014.

\bibitem{Schaft:1999}
A.~van~der Schaft, {\em L2-gain and passivity in nonlinear control}.
\newblock Springer-Verlag, 1999.

\bibitem{Toth2010SpringerBook}
R.~T{\'o}th, {\em Modeling and Identification of Linear Parameter-Varying Systems}.
\newblock Heidelberg: Springer, 2010.

\bibitem{Angeli:1999}
D.~Angeli and E.~Sontag, ``Forward completeness, unboundedness observability, and their lyapunov characterizations,'' {\em Systems \& Control Letters}, vol.~38, no.~4-5, pp.~209--217, 1999.

\bibitem{hjartarson2015lpvtools}
A.~Hjartarson, P.~Seiler, and A.~Packard, ``Lpvtools: A toolbox for modeling, analysis, and synthesis of parameter varying control systems,'' {\em IFAC-PapersOnLine}, vol.~48, no.~26, pp.~139--145, 2015.

\bibitem{Parrilo:2003}
P.~Parrilo, ``Semidefinite programming relaxations for semialgebraic problems,'' {\em Mathematical Programming}, vol.~96, no.~2, pp.~293--320, 2003.

\bibitem{Leung:2017}
K.~Leung and I.~R. Manchester, ``Nonlinear stabilization via control contraction metrics: A pseudospectral approach for computing geodesics,'' in {\em 2017 American Control Conference (ACC)}, pp.~1284--1289, IEEE, 2017.

\bibitem{wang:2019}
R.~Wang and I.~R. Manchester, ``Continuous-time dynamic realization for nonlinear stabilization via control contraction metrics,'' in {\em 2020 American Control Conference (ACC)}, pp.~1619--1624, IEEE, 2020.

\bibitem{Cisneros:2019}
P.~S. Cisneros and H.~Werner, ``Wide range stabilization of a pendubot using quasi-lpv predictive control,'' {\em IFAC-PapersOnLine}, vol.~52, no.~28, pp.~164--169, 2019.

\bibitem{Andrieu2010}
V.~Andrieu and C.~Prieur, ``Uniting two control lyapunov functions for affine systems,'' {\em IEEE Transactions on Automatic Control}, vol.~55, no.~8, pp.~1923--1927, 2010.

\bibitem{Dullerud:2013}
G.~Dullerud and F.~Paganini, {\em A course in robust control theory: {A} convex approach}.
\newblock Springer Science \& Business Media, 2013.

\bibitem{Shamma90c}
J.~Shamma and M.~Athans, ``Analysis of gain scheduled control for nonlinear plants,'' {\em IEEE Transactions on Automatic Control}, vol.~35, no.~8, pp.~898--907, 1990.

\bibitem{Leith98}
D.~Leith and W.~Leithhead, ``Gain-scheduled nonlinear systems: Dynamic analysis by velocity based linearization families,'' {\em International Journal of Control}, vol.~70, pp.~289--317, 1998.

\bibitem{Mehendale2006}
C.~S. Mehendale and K.~M. Grigoriadis, ``Performance of lpv gain-scheduled systems,'' in {\em 2006 American Control Conference}, pp.~6--pp, IEEE, 2006.

\bibitem{shamma1991guaranteed}
J.~S. Shamma and M.~Athans, ``Guaranteed properties of gain scheduled control for linear parameter-varying plants,'' {\em Automatica}, vol.~27, no.~3, pp.~559--564, 1991.

\bibitem{Rotondoi:2019}
D.~Rotondo and M.~Witczak, ``Analysis and design of quadratically bounded qpv control systems,'' {\em IFAC-PapersOnLine}, vol.~52, no.~28, pp.~76--81, 2019.

\bibitem{Sala:2019}
A.~Sala, C.~Ari{\~n}o, and R.~Robles, ``Gain-scheduled control via convex nonlinear parameter varying models,'' {\em IFAC-PapersOnLine}, vol.~52, no.~28, pp.~70--75, 2019.

\bibitem{Koelewijn:2019}
P.~Koelewijn, R.~T{\'o}th, and H.~Nijmeijer, ``Linear parameter-varying control of nonlinear systems based on incremental stability,'' {\em IFAC-PapersOnLine}, vol.~52, no.~28, pp.~38--43, 2019.

\end{thebibliography}

\end{document}